\documentclass{pasj00}
\draft

\begin{document}
\SetRunningHead{Noriaki TAWA}{Reproducibility of NXB for Suzaku XIS}
%\Received{2000/12/31}%{yyyy/mm/dd}
%\Accepted{2001/01/01}%{yyyy/mm/dd}

\title{Reproducibility of Non-X-ray Background for 
the X-ray Imaging Spectrometer aboard Suzaku}

 \author{
   Noriaki \textsc{tawa}\altaffilmark{1},
   Kiyoshi \textsc{hayashida}\altaffilmark{1},
   Masaaki \textsc{nagai}\altaffilmark{1},
   Hajime \textsc{nakamoto}\altaffilmark{1},\\
   Hiroshi \textsc{tsunemi}\altaffilmark{1},
   Hiroya \textsc{yamaguchi}\altaffilmark{2},
   Yoshitaka \textsc{ishisaki}\altaffilmark{3},
   Eric \textsc{miller}\altaffilmark{4}, \\
   Tsunefumi \textsc{mizuno}\altaffilmark{5},
   Tadayasu \textsc{dotani}\altaffilmark{6},
   Masanobu \textsc{ozaki}\altaffilmark{6}, and 
   Haruyoshi \textsc{katayama}\altaffilmark{7}}
 \altaffiltext{1}{Department of Earth and Space Science, Graduate School of Science, Osaka University, Toyonaka, Osaka 560-0043}
 \email{tawa@ess.sci.osaka-u.ac.jp}
 \altaffiltext{2}{Department of Physics, Graduate School of Science, Kyoto University, Sakyo-ku, Kyoto 606-8502}
 \altaffiltext{3}{Department of Physics, Tokyo Metropolitan University, 1-1 Minami-Osawa, Hachioji, Tokyo 192-0397}
 \altaffiltext{4}{Kavli Institute for Astrophysics and Space Research, Massachusetts Institute of Technology, Cambridge, MA 02139, USA}
 \altaffiltext{5}{Department of Physical science, Hiroshima University, 1-3-1 Kagamiyama, Higashi-Hiroshima, Hiroshima 739-8526}
 \altaffiltext{6}{Institute of Space and Astronautical Science, Japan Aerospace Exploration Agency, 3-1-1 Yoshino-dai, Sagamihara, Kanagawa 229-8510}
 \altaffiltext{7}{Japan Aerospace Exploration Agency (JAXA), 2-1-1 Sengen, Tsukuba, Ibaraki 305-8505}
\KeyWords{instrumentation: detectors --- methods: data analysis --- X-rays: general} 

\maketitle

\begin{abstract}
One of the advantages of the X-ray Imaging Spectrometer (XIS) system on board
Suzaku is its low and stable non-X-ray background (NXB). 
In order to make the best use of this advantage, modeling the NXB
spectra with high accuracy is important to subtract them
from the spectra of on-source observations.  
We construct an NXB database by collecting XIS events when 
the dark Earth covers the XIS FOV. The total exposure time of the NXB data 
is about 785\,ks for each XIS. 
It is found that the count rate of the NXB anti-correlates with 
the cut-off-rigidity and correlates with the count rate of the PIN upper 
discriminator (PIN-UD) in Hard X-ray Detector on board Suzaku. 
We thus model the NXB spectrum for a given on-source observation 
by employing either of these parameters and 
obtain a better reproducibility of the NXB for the model with PIN-UD than 
that with the cut-off-rigidity. 
The reproducibility of the NXB model with 
PIN-UD is 4.55-5.63\,\% for each XIS NXB
in the 1-7\,keV band and 
2.79-4.36\,\% for each XIS NXB in 
the 5-12\,keV band for each 5\,ks exposure of the NXB data. 
This NXB reproducibility is much smaller than 
the spatial fluctuation of the cosmic 
X-ray background in the 1-7\,keV band, 
and is almost comparable to that in the 5-12\,keV band.
\end{abstract}

\section{Introduction}\label{intro}

The X-ray Imaging Spectrometer (XIS) at the foci of the
four X-ray telescopes (XRT) on board the Suzaku observatory
are best suited in the recent X-ray astronomy satellites such as
ASCA (\cite{burke}), Chandra (\cite{chandra}), and XMM (\cite{lumb})
for diffuse and low surface brightness sources owing to
their large collection area and low and stable background level.
One of the XIS, XIS1, has a back-illuminated (BI) CCD,
while the other three XIS, XIS0, 2, 3, 
are equipped with front-illuminated (FI) CCDs 
(hereafter referred to as XIS-FIs; \cite{koyama}).
The background levels normalized by the effective area
and the field of view (FOV), in terms of the S/N ratio to diffuse emissions, 
of the XIS-FIs are comparable to those of the ASCA SIS (\cite{mitsuda}).
Moreover, these background levels are $\sim 3$ and $\sim 10$ times
lower than those of the XMM EPIC and the Chandra ACIS at 5\,keV, 
respectively (\cite{mitsuda}).

The background of the XIS consists of three components:
(1) non-X-ray background (NXB); 
(2) a solar component, which is emission from the earth's atmosphere
illuminated by the Sun and solar wind charge exchange; 
and (3) a diffuse X-ray component from such sources a 
local hot bubble (LHB),
galactic diffuse X-ray emissions, and 
cosmic X-ray background (CXB).

While the X-ray background is produced by emission within the XRT FOV, 
the XIS NXB is caused by charged particles and $\gamma$-rays 
(Mizuno et al. 2004) entering the detector from various directions.
Therefore, the NXB varies with time according to 
the radiation environment of the satellite, 
i.e. the particle or $\gamma$-ray spectra hitting on Suzaku.
Since the altitude of orbit of Suzaku is lower than that of XMM or
Chandra, the particle and $\gamma$-ray spectra 
of Suzaku are different from that of XMM or Chandra.
Although these spectra are not entirely clear, 
their intensities of Suzaku are relatively lower and 
more stable than those of XMM or Chandra.
In the case of XMM EPIC, 
solar soft protons produce flares of up to 10 times of the quiescent 
background level and affect 30-40\,\% of XMM observation
time (Carter et al. 2007).
However, this component hardly affects Suzaku.

The solar component (2) includes the fluorescence lines of nitrogen and
oxygen from the earth atmosphere and scattered solar X-rays.
The intensity of the solar component depends both on the solar activity and 
on the elevation angle from the sunlit earth edge.
This component can be minimized by filtering the data based on 
the elevation angle from the sunlit earth edge. 
Emission lines 
of nitrogen and oxygen are enhanced through charge exchange 
between interplanetary and geocoronal neutral atoms and metal ions in 
the solar wind (\cite{fujimoto}).
Referring to the solar wind data, proton or X-ray fluxes, 
helps the evaluation of this component.

The spectrum of the CXB is approximated a power-law spectrum of photon 
index of $\sim 1.4$ in 2-10\,keV band (\cite{kushino}).
The CXB is known to be a collection of faint unresolved extragalactic
sources (\cite{hasinger}). It is uniform over the sky with some fluctuation.
The LHB and galactic diffuse X-ray emissions
are dominant below 1\,keV, and 
their spectra depend on the direction of the sky.
Snowden et al. (1997) proposed a model which provides 
their spectra based on the ROSAT all sky survey.

Among the three components of the XIS background, the solar component is
time variable and most difficult to model.
We try to minimize it 
by using the orbital and altitude data, 
which are elevation angles from the day or night Earth edge, 
and the solar wind data.
The diffuse X-ray component is basically stable
and can be evaluated by observation of other fields by Suzaku.
The target of this paper is to properly estimate 
the NXB so that we can use it as a background model.
It is most important to establish 
a method to accurately evaluate the NXB spectra and time variations 
in order to maximize the advantage of the low background level of the XIS.
We thus  construct an XIS NXB database to be used
in the evaluation of the NXB and 
introduce a method to generate the NXB model 
given the intensity of charged particles.
We also examine and confirm the reproducibility of our NXB model.

\section{NXB of the XIS}

\subsection{NXB database}

We constructed the database of the XIS NXB from 
the events collected while Suzaku was pointed toward
the night Earth (NTE).
Under this condition, the diffuse X-ray component (3) is blocked, 
and the solar component (2) does not contaminate.
The criteria with which we selected the NTE events are as follows.
\begin{itemize}
\item Rev0.7 products (\cite{mitsuda}) of which the XIS
mode was normal $5 \times 5$ or $3 \times 3$ mode 
(without burst or window options). 
The events were further filtered with the condition of 
${\rm T\_SAA\_HXD} > 436$\,s, 
where T\_SAA\_HXD means time after 
the passage of the south atlantic anomaly (SAA).
This criterion is used in 
revision 1.2 or 1.3 products (\cite{mitsuda}), 
and this filtering excludes flares in 
the NXB intensity just after Suzaku passed through the SAA.
The events during the telemetry saturation were also excluded.

\item {\it Cleansis} in FTOOLS was applied with the default parameters
to exclude the flickering pixels.

\item The NTE events were extracted for 
Earth elevation angles (ELV) less than $-5^{\circ}$ and 
Earth day-time elevation angles (DYE\_ELV) greater than $100^{\circ}$.
\end{itemize}
Since the XIS was in initial operation during August 2005, 
we collected the NTE events from data observed between September 2005
and May 2006 with the above criteria.
The total exposure time of the NTE data is $\sim 785$\,ks for each XIS.
The NXB database\footnote{The first version of the database 
is accessible via Suzaku web page at \\
ISAS/JAXA 
$\langle$\,http://www.astro.isas.jaxa.jp/suzaku/analysis/xis/nte/\,$\rangle$ 
and \\
GSFC/NASA 
$\langle$\,http://heasarc.gsfc.nasa.gov/docs/suzaku/analysis/xisbgd0.html\,$\rangle$, \\
but the NXB data in this database 
lack T\_SAA\_HXD and telemetry saturation filtering described here, 
leading to a total exposure of 800\,ks.
The second version (the product of this paper) will be released in 
October 2007.}
consists of
the NTE event files and the associated enhanced house keeping (EHK) file,  
in which orbital information is listed with time.
Two associated tools, 
{\it mk\_corsorted\_spec\_v1.0.pl} and {\it mk\_corweighted\_bgd\_v1.1.pl},
 to generate the NXB model using 
the cut-off-rigidity were also prepared.
Since the event files in the database can be processed with 
various FTOOLS including XSELECT, the NXB spectra can be easily created. 
The subject of this paper is to generate the most appropriate 
NXB spectra for a given observation.

We hereafter refer to the ``NTE events'' as the NXB events and 
refer to the data comprising the NXB events as the ``NXB data''.
Additionally, the ``NXB database'' indicates the data set 
which contains the NXB event
files and the associated EHK file.

\subsection{NXB spectra}

\begin{figure}[htbp]
  \begin{center}
    \FigureFile(80.0mm,59.7mm){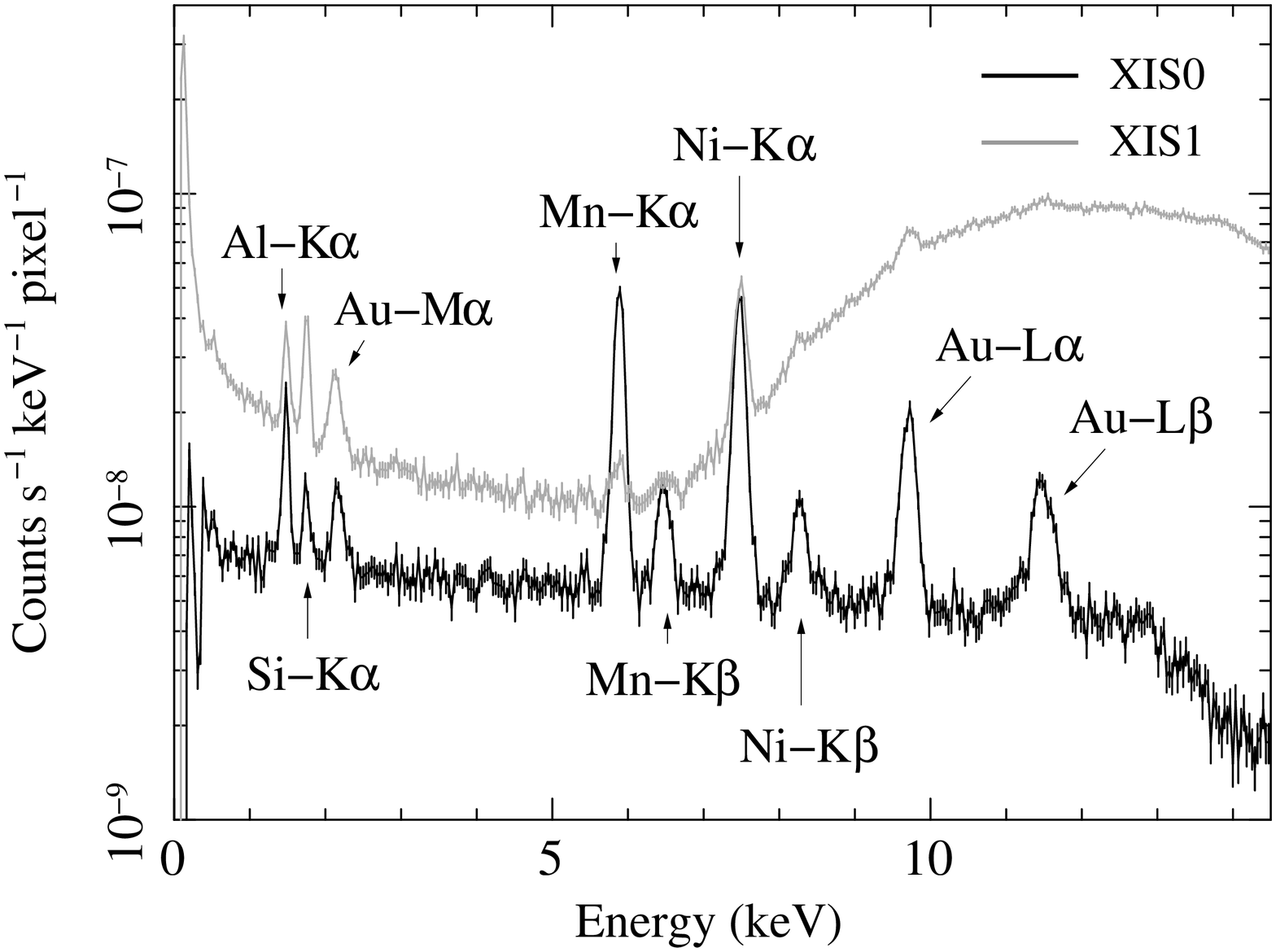}
  \end{center}
    \caption{Spectra of the NXB in the XIS0 (black) and the XIS1 (gray).}
    \label{nxb_spec}
\end{figure}

Figure \ref{nxb_spec} shows the NXB spectra of XIS0 and XIS1.
The spectra are extracted from the whole region of the CCD except for
the calibration source regions (two corners of the CCD chip 
(\cite{koyama})). 
The XIS FOV is $\sim 287$ arcmin$^2$, 
which is 91\,\% of the FOV of the whole CCD chip.
The spectra show fluorescence lines of Al, Si, Au, Mn, and Ni 
in the XIS and XRT.
Table \ref{lines} shows the intensities of these emission lines, 
and table \ref{origin} shows the origin for each fluorescence line.
The XIS0 has relatively strong
Mn-K emission lines at 5.9 and 6.5\,keV as shown in table \ref{lines}.
This is due to stray X-rays from the $^{55}$Fe calibration source, 
although why the radiation is detected outside 
the calibration source regions remains unknown (\cite{yamaguchi_spie}).
Since the XIS-FIs have a thick neutral layer beneath the 
depletion layer, most of the background events generated 
by charged particles produce charge over many pixels and are
rejected as ASCA grade 7 
events\footnote{The ASCA grade shows the spread of an event.
In the case of the XIS, we consider that most of the X-ray events
do not split into a region larger than $2 \times 2$ pixels.
Grade 7 events, in which the spread of event contains more than 
$2 \times 2$ pixels, are regarded as background events (\cite{koyama}).}.
On the other hand, the XIS1 (BI-CCD) has a relatively thin depletion layer
and almost no neutral layer, resulting in 
relatively many background events in grades 0, 2, 3, 4, 
and 6 (\cite{yamaguchi_spie}).
Therefore, the background count rate of the XIS1 is higher than those of the
XIS-FIs, especially above $\sim 7$\,keV as shown in Fig. \ref{nxb_spec}.

\begin{table*}[htbp]
  \caption{Energies and count rates of the line components in the NXB spectra.}
  \label{lines}
  \begin{center}
    \begin{tabular}{l c c c c c}
      \hline \hline
Line & Energy & \multicolumn{4}{c}{Count rate\footnotemark[$*$] ($10^{-9}$ cts s$^{-1}$ pixel$^{-1}$)} \\
\cline{3-6} 
 & (keV) & XIS0 & XIS1 & XIS2 & XIS3  \\
      \hline
Al-K$\alpha$ & 1.486 & $1.45 \pm 0.11$ & $1.84 \pm 0.14$ & $1.41 \pm 0.10$ & $1.41 \pm 0.10$ \\
Si-K$\alpha$ & 1.740 & $0.479 \pm 0.081$ & $2.27 \pm 0.15$ & $0.476 \pm 0.080$ & $0.497 \pm 0.082$ \\
Au-M$\alpha$ & 2.123 & $0.63 \pm 0.093$ & $1.10 \pm 0.13$ & $0.776 \pm 0.097$ & $0.619 \pm 0.092$ \\
Mn-K$\alpha$ & 5.895 & $6.92 \pm 0.19$ & $0.43 \pm 0.14$ & $1.19 \pm 0.13$ & $0.76 \pm 0.11$ \\
Mn-K$\beta$ & 6.490 & $1.10 \pm 0.11$ & $0.26 \pm 0.13$ & $0.40 \pm 0.11$ & $0.253 \pm 0.094$ \\
Ni-K$\alpha$ & 7.470 & $7.12 \pm 0.19$ & $7.06 \pm 0.37$ & $8.01 \pm 0.20$ & $7.50 \pm 0.20$ \\
Ni-K$\beta$ & 8.265 & $0.96 \pm 0.10$ & $0.75 \pm 0.22$ & $1.16 \pm 0.11$ & $1.18 \pm 0.11$ \\
Au-L$\alpha$ & 9.671 & $3.42 \pm 0.15$ & $4.15 \pm 0.49$ & $3.45 \pm 0.15$ & $3.30 \pm 0.15$ \\
Au-L$\beta$ & 11.51 & $2.04 \pm 0.14$ & $1.93 \pm 0.48$ & $1.97 \pm 0.14$ & $1.83 \pm 0.14$ \\
      \hline
      \multicolumn{6}{@{}l@{}}{\hbox to 0pt{\parbox{180mm}{\footnotesize
\par\noindent 
\footnotemark[$*$] The count rates are obtained from the whole CCD chip 
excluding the calibration source \\[-5pt]
regions. Errors are 90\,\% confidence level.
}\hss}}
    \end{tabular}
  \end{center}
\end{table*}

\begin{table}[htbp]
  \caption{Origins of the fluorescence lines in the NXB spectra.}
  \label{origin}
  \begin{center}
    \begin{tabular}{l l}
      \hline \hline
Line & \multicolumn{1}{c}{Origin} \\
      \hline
Al-K$\alpha$ & Optical blocking filter, housing, 
alumina substrate to mount CCD \\
Si-K$\alpha$ & CCD (Si fluorescence line) \\
Au-M$\alpha$, L$\alpha$, L$\beta$ & Housing, CCD substrate, heatsink \\
Mn-K$\alpha$, K$\beta$ & Scattered X-rays from calibration sources\\
Ni-K$\alpha$, K$\beta$ & Housing, heatsink \\
      \hline
    \end{tabular}
  \end{center}
\end{table}

\subsection{Cut-off-rigidity and PIN-UD}\label{cor}

Since the NXB is caused by charged particles, 
the NXB should depend on the intensity of charged particles 
striking Suzaku, and this is strongly correlated with the geomagnetic
cut-off-rigidity.
We introduce a new type of cut-off-rigidity, {\it COR2}, for Suzaku.
The calculation for the {\it COR2} is independent from that for 
the conventional cut-off-rigidity, {\it COR}.
The {\it COR} has been employed 
in the analysis of Tenma, Ginga, ASCA, and Suzaku.
We summarize the characteristics of {\it COR} and {\it COR2} in 
appendix \ref{oldCOR}. 
In the main text, we use {\it COR2} when 
discussing the NXB.

\begin{figure}
   \begin{center}
      \FigureFile(80.0mm,59.0mm){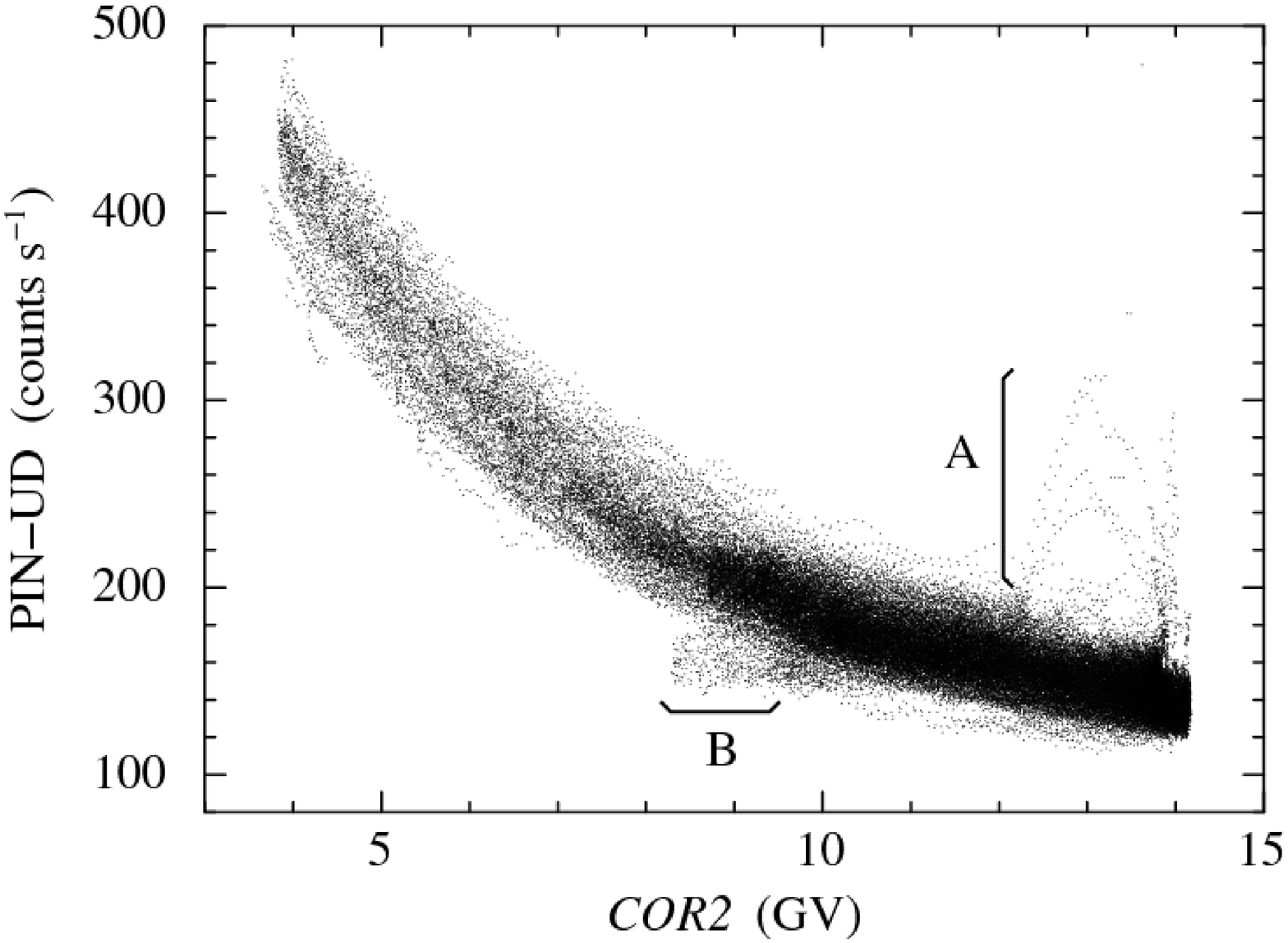}
   \end{center}
   \caption{The PIN-UD as a function of the {\it COR2}.
The PIN-UD is anti-correlated with the {\it COR2}, 
although there is noticeable scatter such as the region of A and B.
The events in region A occurred just before Suzaku entered the SAA.
The events in region B occurred at the time when 
Suzaku passed near the region of 
$({\rm longitude,\ latitude})=(350^{\circ},\ 19^{\circ})$.}
\label{pinud_cor}
\end{figure}

Suzaku carries a non-imaging hard X-ray instrument, the Hard X-ray
Detector (HXD).
The HXD sensor contains $4 \times 4$ well-type phoswich units (well units)
with 4 PIN silicon diodes in each (\cite{hxd}, \cite{kokubun}).
When a charged particle
generates a large signal in a PIN silicon diode, the PIN upper discriminator 
(PIN-UD) is
activated at a threshold around 90\,keV.
This can be a good monitor of the real-time intensity of the charged
particles striking Suzaku. 
The number of PIN-UD counts is recorded with each well unit.
We sum up the PIN-UD count rates for all well units and 
average them for each 32 seconds 
to reduce the statistical error.
The typical number of PIN-UD counts in 32 seconds is $\sim 5100$ counts
by summing up all well units.
We hereafter call this count rate the PIN-UD.
Figure \ref{pinud_cor} shows the PIN-UD as a function of the {\it COR2}. 
There is a strong
anti-correlation between the PIN-UD and the {\it COR2}.
However, the anti-correlation is widely distributed.
This is because that the PIN-UD mirrors
the real-time intensity of the charged particles, while the {\it COR2}
is calculated from a COR map (shown in Fig. \ref{cor_map}(b)) and 
the orbital position of Suzaku.
Therefore, the {\it COR2} might not correctly reproduce 
the real-time intensity of the charged particles.
In addition, some events
deviate from the correlation as shown in the region of A 
and B of Fig. \ref{pinud_cor}. 
The events in region A 
occurred just before Suzaku entered the SAA.
The events in region B occurred at the time when 
Suzaku passed near the region of 
$({\rm longitude,\ latitude})=(350^{\circ},\ 19^{\circ})$.
We assume that this is because the {\it COR2} values in this region
are approximately calculated as shown in appendix 1.
However, the number of events in regions A and B is less than 
1\,\% of the total number of events.
We will discuss which parameters of the {\it COR2} or PIN-UD 
can correctly reproduce the NXB in the next section.

\begin{figure}[htbp]
  \begin{center}
    \FigureFile(160.0mm,59.1mm){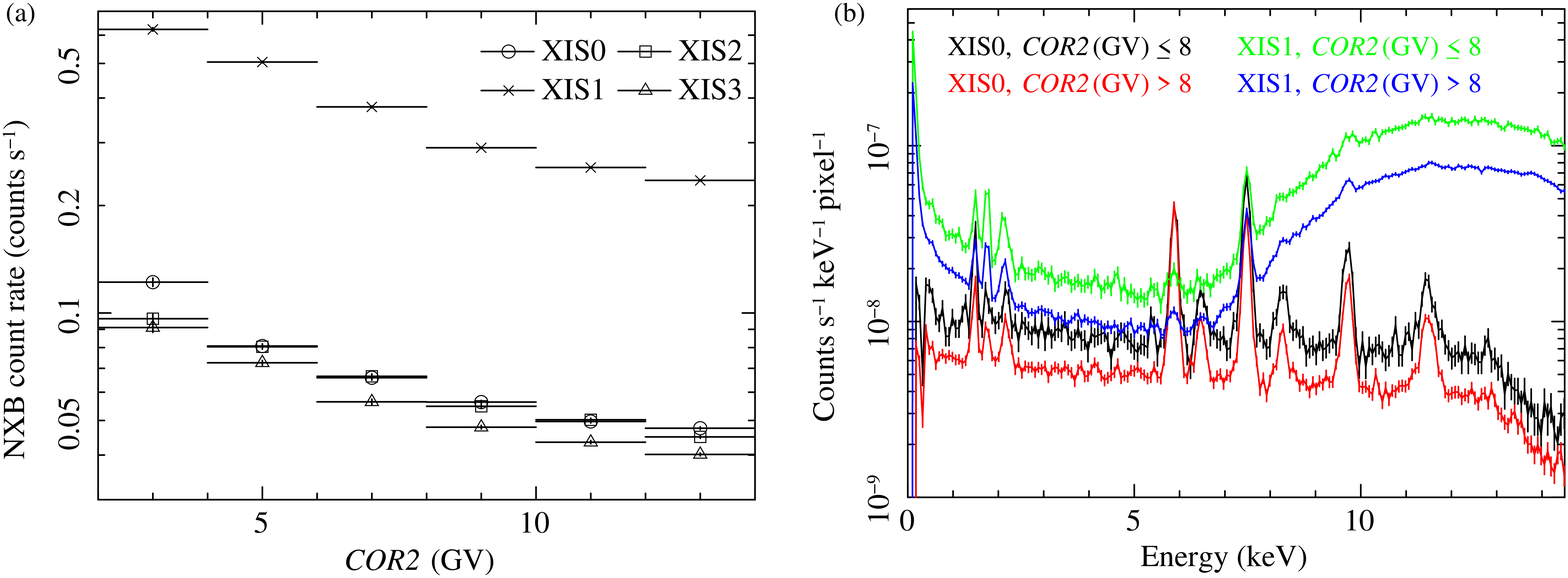}
  \end{center}
  \caption{(a) {\it COR2} dependence of the NXB 
(average count rate in 5-12 keV) for each XIS. 
(b) NXB spectra in ${\it COR2} \le 8$ 
and ${\it COR2} > 8$ GV. 
Black and red lines show the XIS0 spectra.
Green and blue lines show the XIS1 spectra.
The NXB count rate is anti-correlated with the {\it COR2}, and
the count rate of XIS1 is higher than those of XIS-FIs.}
  \label{nxb_cps}
\end{figure}

Figure \ref{nxb_cps}(a) shows the count rate of the NXB 
for each XIS in the 5-12\,keV
energy band as a function of the {\it COR2}. 
The count rate of each bin of XIS1 is 
about 6 times higher than those of XIS-FIs.
This is because the XIS1 has a relatively thin depletion layer
and almost no neutral layer, as discussed in section 2.2.
Therefore, 
the NXB intensity of XIS1 depends on the intensity of 
charged particles as well as the XIS-FIs.
On the other hand, 
spectra are different between the low {\it COR2} region (${\it COR2} \le 8$ GV)
and high {\it COR2} region (${\it COR2} > 8$ GV) 
as shown in Figure \ref{nxb_cps}(b). 
The differences mainly appear in normalization of the spectra.
The periods of the NXB variations primarily 
correspond to the orbital period of Suzaku, 96 minutes, 
since the NXB depends on the cut-off rigidity.
We should also note that the NXB does not have
apparent long-term changes in 9 months. 
For details, we will discuss in sections 3.3 and 4.

Since the {\it COR2} and the NXB count rate are anti-correlated, 
we can use the {\it COR2} to 
estimate the NXB spectra to be subtracted 
as background for a given observation.
The PIN-UD can also be used as such 
a parameter to estimate the NXB spectra, 
considering the anti-correlation between the {\it COR2} and the PIN-UD. 
In the following section, we will attempt to model the 
NXB spectra from the NXB data by employing either of the {\it COR2} or 
the PIN-UD. The two kinds of NXB models, one with the 
{\it COR2} and the other with the PIN-UD, will be compared by
their reproducibilities.

%%%%%%%%%%%%%%%%%%%%%%%%%%%%%%%%%%%%%%%%%%%%%%%%%%%%%%%%%%%%%%
\section{NXB models and their Reproducibility}\label{sct_repro}

\subsection{NXB models: NXB data sorted by {\it COR2} or PIN-UD}\label{howto_model}

In order to correctly subtract the NXB 
from on-source observation, we introduce method to model the NXB.
Since the spectra of charged particles and $\gamma$-ray
causing the NXB are not fully understood, 
this is a semi-empirical method.
We prepared two tools, 
{\it mk\_corsorted\_spec} and {\it mk\_corweighted\_bgd}\,\footnote{These 
tools are new versions of 
{\it mk\_corsorted\_spec\_v1.0.pl} and {\it mk\_corweighted\_bgd\_v1.1.pl}.\\
{\it mk\_corsorted\_spec\_v1.0.pl} and {\it mk\_corweighted\_bgd\_v1.1.pl}
support the {\it COR} only.
We will merge the new tools in one
and release it as {\it xisnxbgen} in FTOOLS.}, 
to generate the NXB model.
{\it Mk\_corsorted\_spec} is to sort with modeling parameter, 
i.e. {\it COR2} or PIN-UD, bin and generates
the NXB spectra for each modeling parameter bin.
{\it Mk\_corweighted\_bgd} is to generate the NXB model spectrum for a given
on-source observation by summing up the sorted spectra with 
appropriate weights.
The weighted NXB model spectrum, $S_w$, is expressed as follows,
\begin{eqnarray}
S_w = \frac{\sum^n_{i=1} T_i S_i}{\sum^n_{i=1} T_i} =\sum^n_{i=1} 
\frac{T_i}{T_{total}} S_i \ ,
\label{eq2}
\end{eqnarray}
where the modeling parameter is sorted into $n$ bins. 
$T_i$ and $S_i$ are the exposure time of the on-source observation and 
the spectrum of the NXB data in the $i$th modeling parameter bin, respectively.
$T_{total}$ is the total exposure time of the on-source observation.
Equation (\ref{eq2}) makes equal the modeling parameter 
distribution for the on-source
observation and that for the NXB data.

We sorted the NXB and on-source data into 14 bins with either the {\it COR2} 
or PIN-UD. 
The {\it COR2} and PIN-UD bins are defined as shown in table \ref{range}.
In addition, table \ref{range} shows the NXB count rate (5-12\,keV)
and exposure time for each {\it COR2} and PIN-UD bin in the XIS0.
We defined the bin ranges at even intervals of the PIN-UD and
set the {\it COR2} bins so as to get the approximately comparable
count rate of the corresponding PIN-UD bins.
We should note that the NXB model obtained by sorting into 
even intervals with the exposure time 
has comparable level to 
that by sorting into the bins shown in table \ref{range}.

\begin{table}[htbp]
  \caption{The PIN-UD and the {\it COR2} bins ranges.}\label{range}
  \begin{center}
    \begin{tabular}{l c c c c c c c c c c}
      \hline \hline
      Bin \# & PIN-UD & Count rate\footnotemark[$*$] & Exposure\footnotemark[$\dagger$] & & {\it COR2} & Count rate\footnotemark[$*$] & Exposure\footnotemark[$\dagger$] \\
      \cline{2-4} \cline{6-8} 
      & (cts s$^{-1}$) & $10^{-2}$ (cts s$^{-1}$) & (ks) & & (GV) & $10^{-2}$ (cts s$^{-1}$) & (ks) \\
      \hline
      1 & 100-150 & $4.551 \pm 0.051$ & 175.4 & & 15-12.8 
& $4.746 \pm 0.048$ & 208.3 \\
      2 & 150-175 & $4.857 \pm 0.045$ & 236.6 & & 12.8-10.5 
& $4.877 \pm 0.045$ & 246.2 \\
      3 & 175-200 & $5.329 \pm 0.062$ & 137.7 & & 10.5-9.1 
& $5.347 \pm 0.061$ & 142.6 \\
      4 & 200-225 & $5.533 \pm 0.089$ & 70.0 & & 9.1-8.1 
& $5.73 \pm 0.10$ & 53.6 \\
      5 & 225-250 & $6.24 \pm 0.15$ & 28.3 & & 8.1-7.3 
& $6.24 \pm 0.15$ & 27.6 \\ 
      6 & 250-275 & $6.74 \pm 0.18$ & 20.7 & & 7.3-6.6 
& $6.49 \pm 0.18$ & 19.2 \\
      7 & 275-300 & $6.85 \pm 0.20$ & 17.9 & & 6.6-6.0 
& $7.17 \pm 0.20$ & 17.5 \\
 8 & 300-325 & $7.83 \pm 0.23$ & 14.3 & & 6.0-5.5 
& $7.29 \pm 0.24$ & 13.0 \\
 9 & 325-350 & $7.42 \pm 0.23$ & 13.6 & & 5.5-5.1 
& $7.83 \pm 0.27$ & 10.8 \\
 10 & 350-375 & $8.35 \pm 0.26$ & 12.5 & & 5.1-4.7 
& $7.52 \pm 0.25$ & 12.2\\
 11 & 375-400 & $9.33 \pm 0.27$ & 12.4 & & 4.7-4.3 
& $8.60 \pm 0.28$ & 11.3 \\
 12 & 400-425 & $10.61 \pm 0.31$ & 10.8 & & 4.3-4.0 
& $9.55 \pm 0.32$ & 9.5 \\
 13 & 425-450 & $11.53 \pm 0.48$ & 4.9 & & 4.0-3.7 & $11.37 \pm 0.39$ & 7.3 \\
 14 & 450-500 & $7.4 \pm 1.3$ & 0.4 & & 3.7-2.0 & $13.74 \pm 0.59$ & 4.0 \\
      \hline
      \multicolumn{6}{@{}l@{}}{\hbox to 0pt{\parbox{180mm}{\footnotesize
\par\noindent \\[-6pt] 
\footnotemark[$*$] The count rates are obtained from the NXB data 
of the XIS0 in the 5-12\,keV band. Errors are $1 \sigma$ 
confidence \\[-5pt] level.\\[-5pt]
\footnotemark[$\dagger$] The exposure times are obtained from the NXB data 
of the XIS0.
}\hss}}
    \end{tabular}
  \end{center}
\end{table}

\subsection{Reproducibility of the NXB models}\label{sec_repro}

\begin{figure}[htbp]
  \begin{center}
    \FigureFile(160.0mm,63.1mm){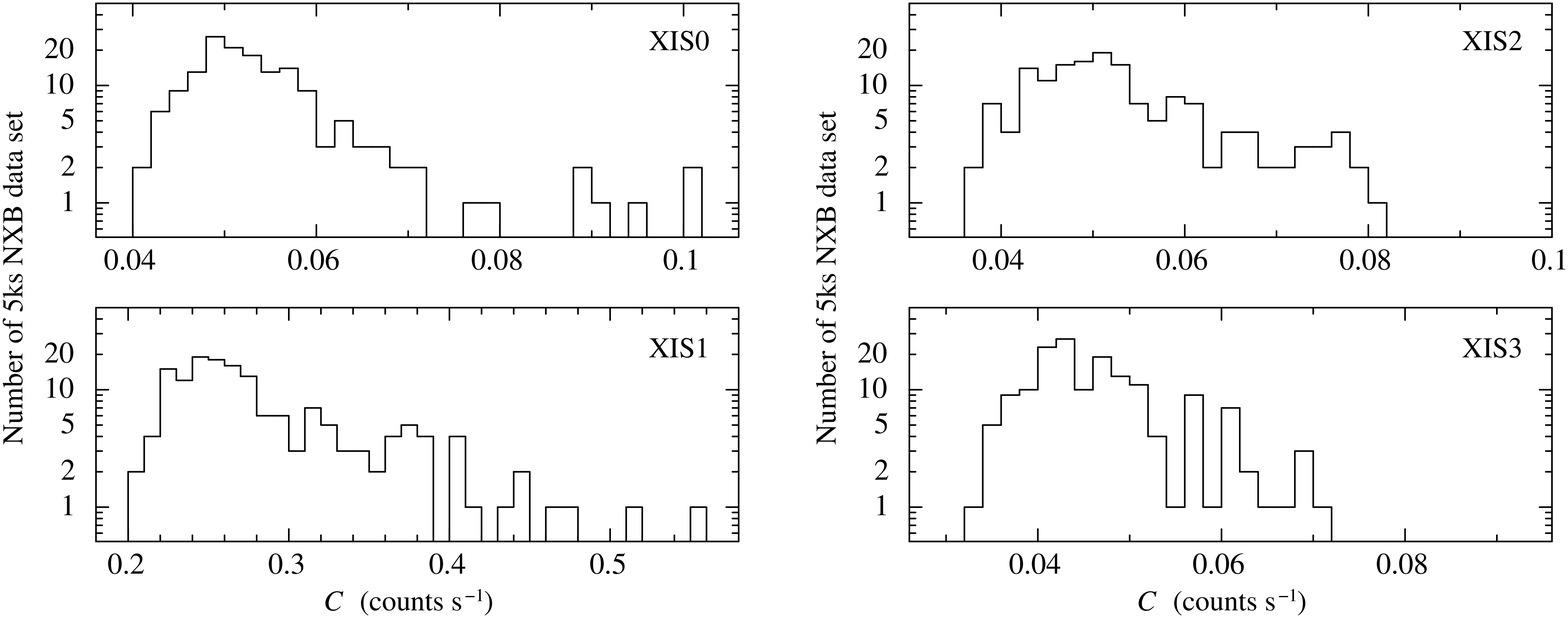}
  \end{center}
  \caption{Distribution of $C_j$ in the XIS0, 1, 2, and 3.
$C_j$ is calculated in the 5-12\,keV energy band.}
    \label{distri_org}
\end{figure}

We first calculate the intrinsic variability of the NXB data to compare with 
the reproducibility of the NXB model.
The standard deviation of the NXB count rate contains 
the systematic error and statistical error.
We define the systematic error ($1 \sigma$ confidence level) 
as the intrinsic variability.
To calculate the intrinsic variability, 
we divide the NXB data into 5\,ks exposure bins 
(generally spanning a few days) 
and obtain the count rate for each.
Since the NXB intensity is particularly low, 
the count rates are obtained in the 5-12\,keV energy band 
to reduce the statistical error.
There are typically 250 counts
per 5\,ks exposure bin in this energy range for the XIS-FIs.
Figure \ref{distri_org} shows the distribution of 
this count rate, hereafter called $C_j$ for the $j$th bin.
We calculate the standard deviation of $C_j$ ($\sigma_c$) as follows,
\begin{eqnarray}
\sigma_c^2 = \frac{1}{n-1} \sum^n_{j=1} (C_j - \mu_c)^2 \ ,
\label{a2-2}
\end{eqnarray}
where $n$ is the number of the 5\,ks NXB data, and 
$\mu_c$ is the average of $C_i$, $\mu_c = (1/N) \sum^n_{j=1} C_j$.
Then, the statistical error of $C_j$ is assumed by Poisson statistics 
and is calculated by $\sqrt{C_j / T_j}$, where 
$T_j$ is a exposure time of the $j$th bin (in this case, $T_j$ is 5\,ks).
Since we divided the NXB data into 5\,ks exposure,
the statistical error of each $C_j$ is approximately constant.
We therefore obtain the statistical error 
contained in $\sigma_c$ ($\sigma_{sta,c}$) as follows,
\begin{eqnarray}
\sigma_{sta,c} = \frac{1}{n} \sum^n_{j=1} \sqrt{ \frac{C_j}{T_j} } \ .
\label{a2-3}
\end{eqnarray}
We then calculate the systematic error ($\sigma_{sys,c}$) as follows,
\begin{eqnarray}
\sigma_{sys,c} = \sqrt{ \sigma_c^2 - \sigma_{sta,c}^2 } \ .
\label{a2-4}
\end{eqnarray}
$\sigma_{sys,c}$ is the intrinsic variability and is
summarized in table \ref{stat_nxbdata}.
The intrinsic variability shows the reproducibility of the NXB 
without being modeled.
Details about the errors of $\sigma_c$, $\sigma_{sta,c}$,
and $\sigma_{sys,c}$ shown in table \ref{stat_nxbdata},
are presented in appendix 2.

\begin{figure}[htbp]
  \begin{center}
    \FigureFile(160.0mm,125.4mm){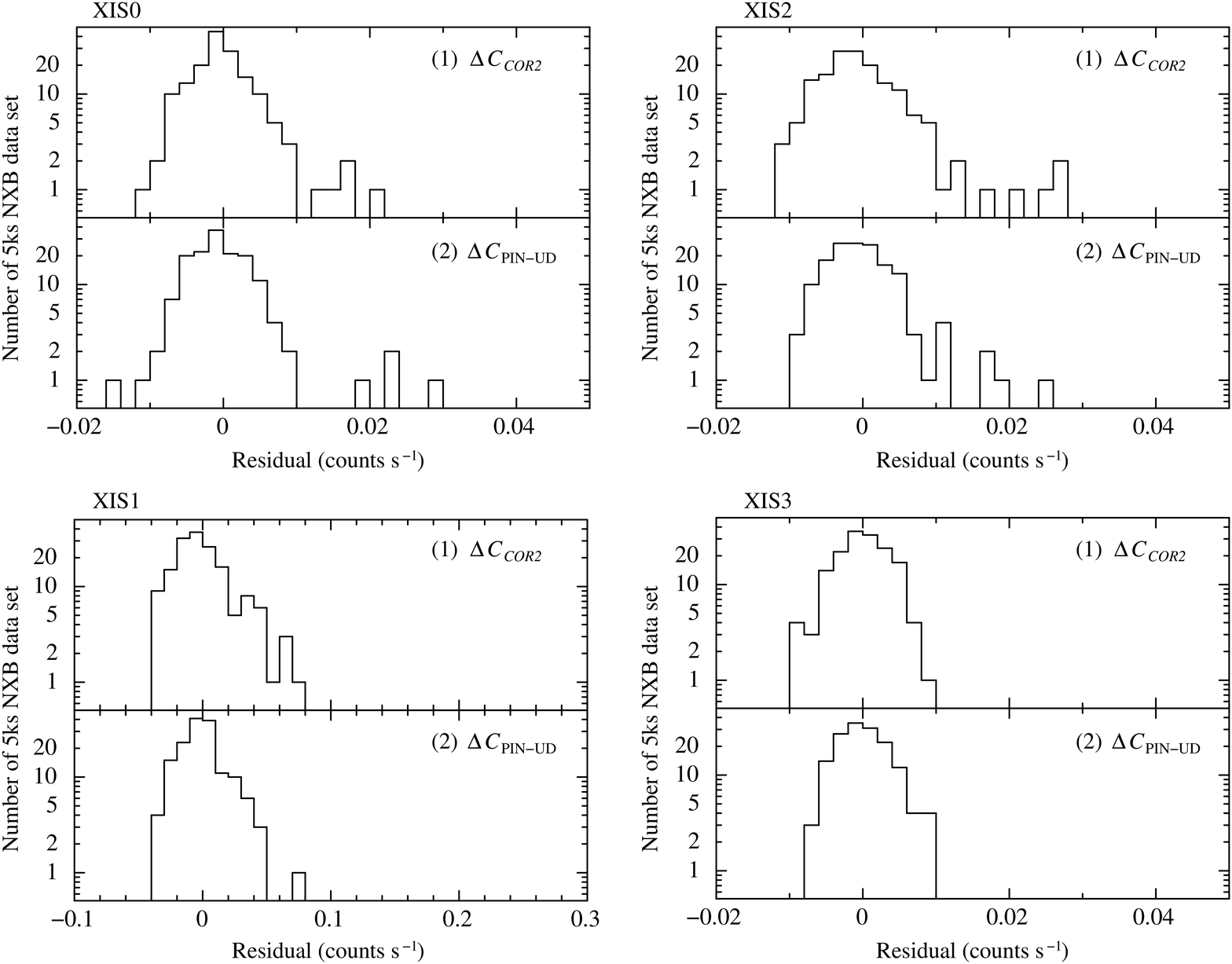}
  \end{center}
  \caption{Distributions of (1) $\Delta C_{COR2}$, 
and (2) $\Delta C_{{\rm PIN-UD}}$ in 
the XIS0, 1, 2, and 3.}
    \label{distri}
\end{figure}

We next calculate the reproducibility of NXB model 
described in eq. (\ref{eq2}).
The NXB spectra, $S_i$ in eq. (\ref{eq2}), are obtained from the 
NXB data according to the modeling parameter.
The weights of each bin, $T_i/T_{total}$ in eq. (\ref{eq2}), 
are calculated by the modeling parameter for each 5\,ks NXB data bin.
We thus obtained the NXB models for each 5\,ks bin and 
calculated the residual, data minus model.
The $j$th residual ($\Delta C_j$) is calculated by $C_j - M_j$, 
where $M_j$ is the count rate of the NXB model for the $j$th 
5\,ks NXB data bin.
There are two kinds of $\Delta C_j$, 
$\Delta C_{COR2}$ and $\Delta C_{{\rm PIN-UD}}$, which are 
calculated based on the {\it COR2} and the PIN-UD, respectively.
Figure \ref{distri} shows the distributions of 
$\Delta C_{COR2}$ and $\Delta C_{{\rm PIN-UD}}$ in the 5-12\,keV energy band.
These distributions are relatively narrow 
compared with the distribution of $C_j$ 
shown in Fig. \ref{distri_org}.
This indicates that the NXB models correctly reproduce the NXB data.
Since the way to calculate the reproducibility is the same with 
$\Delta C_{COR2}$ and $\Delta C_{{\rm PIN-UD}}$, 
we express this with $\Delta C_j$.
The standard deviation of $\Delta C_j$ ($\sigma_{\Delta c}$) 
is calculated as follows,
\begin{eqnarray}
\sigma_{\Delta c}^2 = \frac{1}{n-1} \sum^n_{j=1} (\Delta C_j - 
\mu_{\Delta c})^2 \ ,
\end{eqnarray}
where $\mu_{\Delta c}$ is the average of $\Delta C_j$ and 
is expected to be zero.
The statistical error of $\Delta C_j$ is $\sqrt{C_j/T_j + M_j/T}$, where
$T$ is total exposure time of the NXB data.
The average of these statistical errors ($\sigma_{sta,\Delta c}$) 
is expressed as follows, 
\begin{eqnarray}
\sigma_{sta,\Delta c} = \frac{1}{n} \sum^n_{j=1} 
\sqrt{\frac{C_j}{T_j} + \frac{M_j}{T}} \ . 
\end{eqnarray}
Since $T \sim 157 T_j$, 
the value of $\sigma_{sta,\Delta c}$ is approximately $\sigma_{sta,c}$ 
(eq. (\ref{a2-3})).
By using  $\sigma_{\Delta c}$ and $\sigma_{sta,\Delta c}$, 
the systematic error of $\Delta C_j$ ($\sigma_{sys,\Delta c}$) 
is calculated as follows,
\begin{eqnarray}
\sigma_{sys,\Delta c} = \sqrt{ \sigma_{\Delta c}^2 - 
\sigma_{sta,\Delta c}^2 } \ .
\end{eqnarray}
Hear $\sigma_{sys,\Delta c}$ is defined as 
the ``reproducibility'' of the NXB model.
We independently calculate the reproducibility for each XIS and
show them in table \ref{stat_model}(a).
Details about the errors of the reproducibilities are presented 
in appendix 2.
Since the PIN-UD sometimes exceeds
the range of 100-500\,cts s$^{-1}$,
the total exposure time reduces to $\sim 760$\,ks.
On the other hand, for the NXB model with the {\it COR2}, 
the whole NXB data set of $\sim 785$\,ks is available.
The reproducibilities of the NXB models (table \ref{stat_model}(a))
are about 1/3 of the intrinsic variability of the NXB
count rate (table \ref{stat_nxbdata}).
However, the residuals sometimes becomes large
in both NXB models, as shown in Fig. \ref{distri}.

\begin{table}[htbp]
  \caption{Statistical summary of the NXB data.}
  \label{stat_nxbdata}
  \begin{center}
    \begin{tabular}{l c c c c}
      \hline \hline
Sensor & Average count rate\footnotemark[$*$] & Standard deviation\footnotemark[$\dagger$] & Statistical error\footnotemark[$\dagger$] & Intrinsic variability\footnotemark[$\dagger$] \\
 & $10^{-2}$ (cts s$^{-1}$) & (\%) & (\%) & (\%) \\
      \hline
        XIS0 & $5.464 \pm 0.026$ & $19.2 \pm 1.1$ & $6.03 \pm 0.34$ & $18.2 \pm 1.1$ \\
        XIS1 & $28.758 \pm 0.060$ & $22.5 \pm 1.3$ & $2.62 \pm 0.15$ & $22.4 \pm 1.3$ \\
        XIS2 & $5.317 \pm 0.026$ & $19.1 \pm 1.1$ & $6.11 \pm 0.35$ & $18.1 \pm 1.1$ \\
        XIS3 & $4.685 \pm 0.024$ & $16.85 \pm 0.95$ & $6.51 \pm 0.37$ & $15.5 \pm 1.0$ \\
      \hline
      \multicolumn{5}{@{}l@{}}{\hbox to 0pt{\parbox{180mm}{\footnotesize
\par\noindent 
\footnotemark[$*$] The count rates are obtained from the NXB data 
in the 5-12\,keV band.  Errors are $1 \sigma$ confidence level.\\[-5pt]
\footnotemark[$\dagger$] These values are normalized by the average count rate.
}\hss}}
    \end{tabular}
  \end{center}
\end{table}

\subsection{Improvements to the NXB model by filtering the data}

\begin{figure}[htbp]
  \begin{center}
    \FigureFile(80.0mm,57.2mm){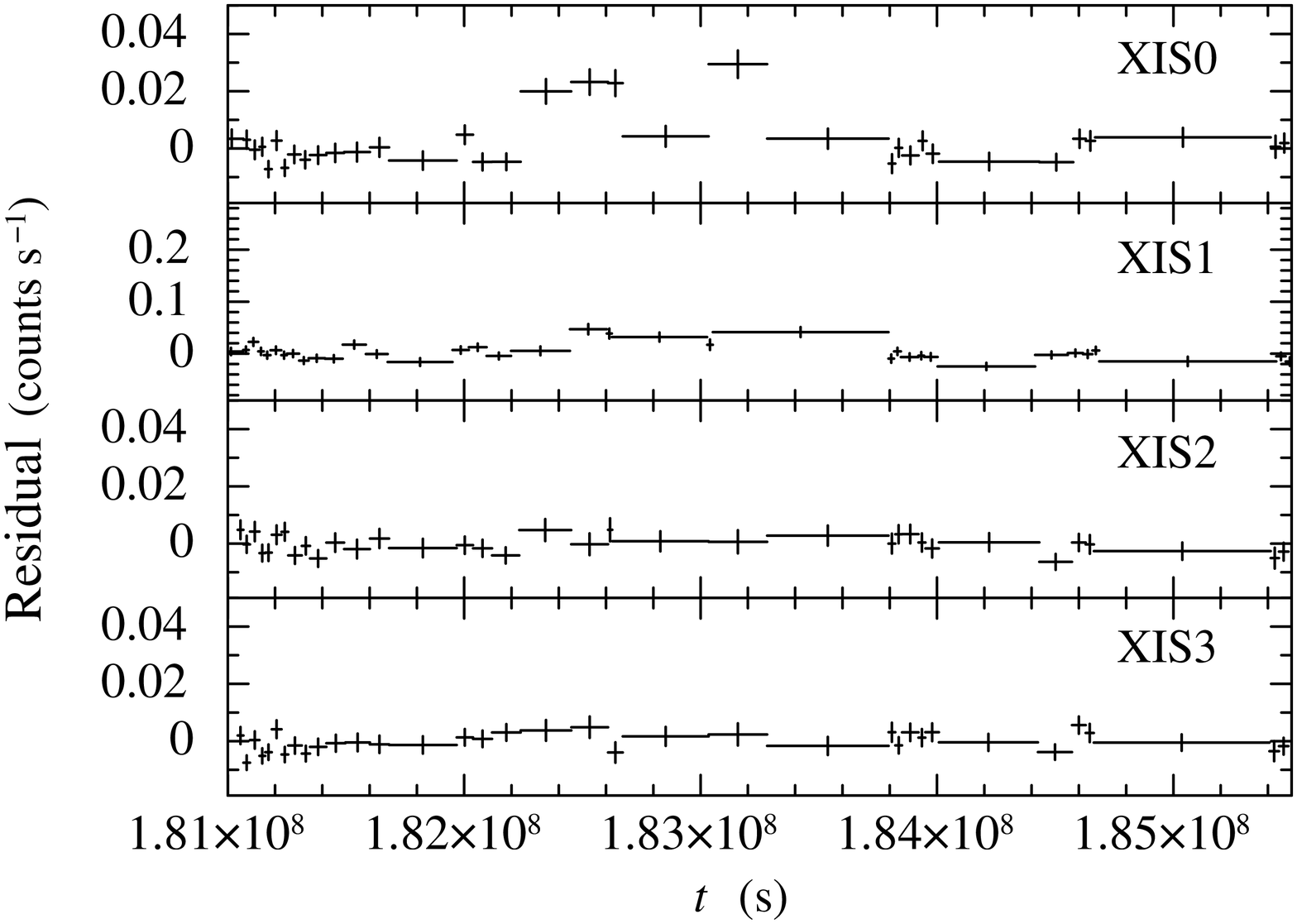}
  \end{center}
  \caption{Light curve of $\Delta C_{{\rm PIN-UD}}$ in the XIS0, 1, 2, and 3.
The residuals are obtained from the unfiltered NXB data in the 5-12keV band.}
    \label{lc_of}
\end{figure}

To further improve the reproducibilities of the NXB models, 
we examined the time and orbital position of Suzaku 
when the count rate significantly deviates from the NXB model,
i.e. $> 0.01$ cts s$^{-1}$ for the XIS-FIs in Fig. \ref{distri}.
We found that those data are obtained from 2005/10/09 
($t = 1.822 \times 10^{8}$ s, where $t$ is time since 2000/01/01 00:00:00) 
to 2005/10/28 
($t = 1.838 \times 10^{8}$ s) and from 2005/11/29 
($t = 1.866 \times 10^{8}$ s) to 2005/12/20
($t = 1.884 \times 10^{8}$ s).
We call these time periods period-A.
Figure \ref{lc_of} shows the light curve of $\Delta C_{PIN-UD}$.
The count rate of $\Delta C_{PIN-UD}$ increase
at the first time period of period-A, especially in the XIS0.

\begin{figure}[htbp]
  \begin{center}
    \FigureFile(80.0mm,58.6mm){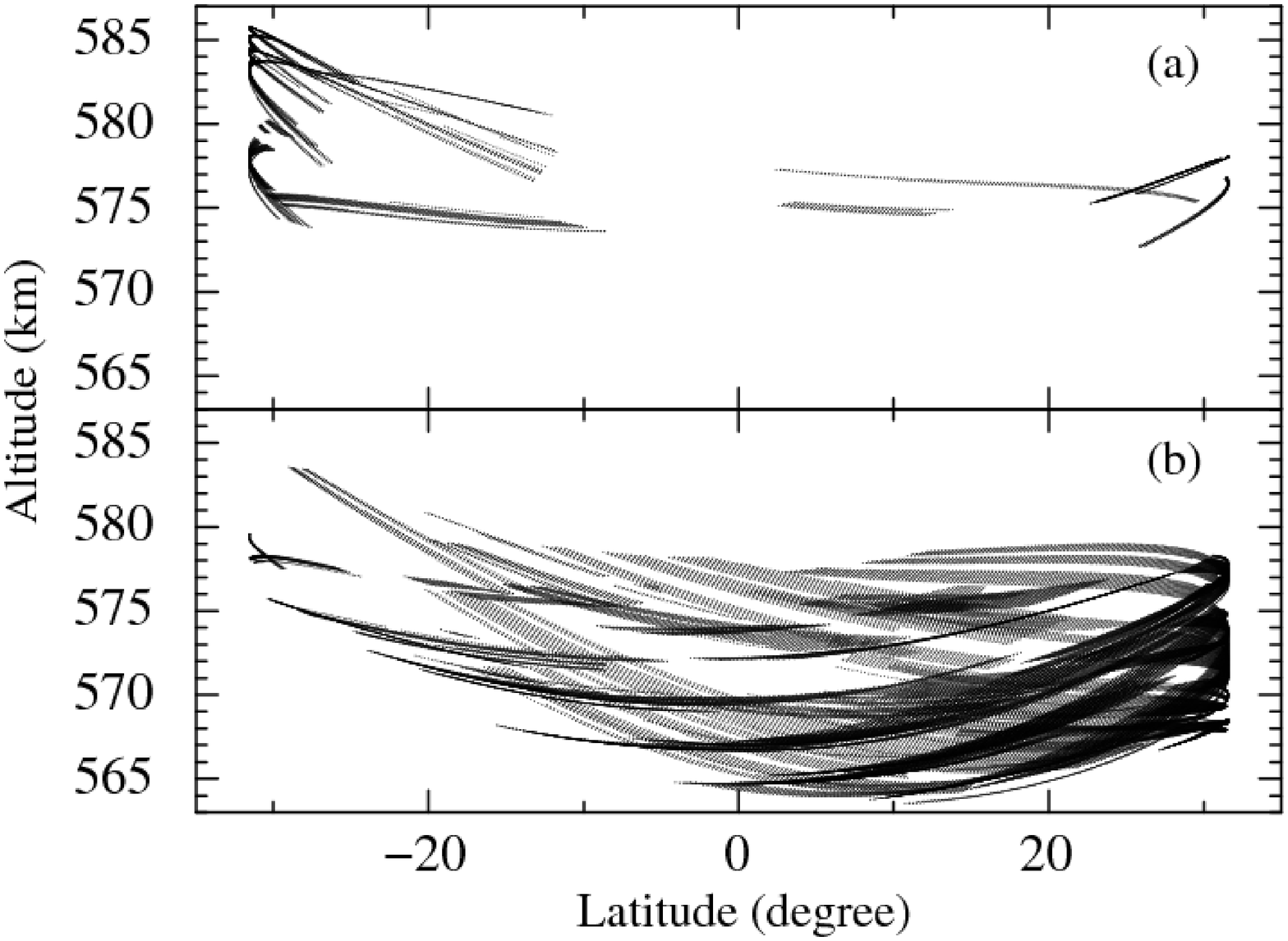}
  \end{center}
  \caption{Orbital positions of Suzaku for XIS observations of the 
NTE during (a) period-A and (b) other times.}
    \label{orbit}
\end{figure}

\begin{figure}[htbp]
  \begin{center}
    \FigureFile(160.0mm,125.4mm){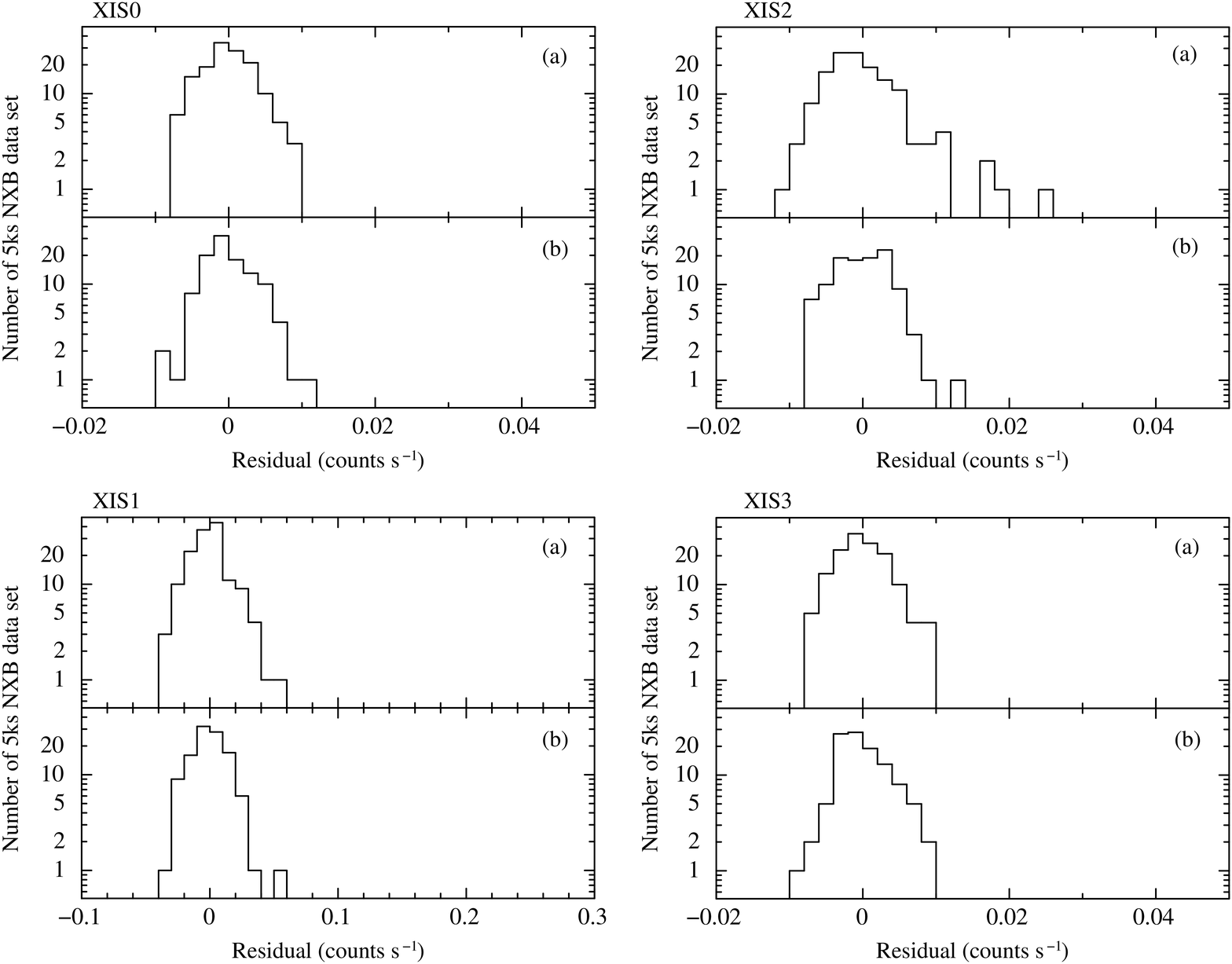}
  \end{center}
  \caption{Distribution of $\Delta C_{{\rm PIN-UD}}$ in 
the XIS0, 1, 2, and 3.
The residuals are obtained from (a) the {\it NXB1} and (b) the {\it NXB2} data
in 5-12\,keV energy band.}
    \label{of_distri}
\end{figure}

Figure \ref{orbit}(a) shows the orbital position during period-A,
and Fig. \ref{orbit}(b) shows the orbital position at other times.
These plots indicate that 
the residuals increase when Suzaku passes through
high latitude and high altitude region.
We therefore exclude the NXB events during the time when 
the orbital positions of Suzaku were
${\rm latitude \le -23^{\circ},\ altitude \ge 576.5\ km}$
or ${\rm latitude \ge 29^{\circ},\ altitude \ge 577.5\ km}$ 
(hereafter the ``orbit filter'') from the NXB data.
The total exposure time of the NXB data with the orbit filter 
(hereafter ``{\it NXB1}'') 
is $\sim 730$\,ks, while that with 
the PIN-UD is $\sim 710$\,ks.
The reproducibilities of the NXB models for the {\it NXB1} data are 
independently evaluated for each XIS by
the same way as that in section \ref{sec_repro} .
Figure \ref{of_distri}(a) shows the distribution 
of $\Delta C_{{\rm PIN-UD}}$ obtained from the {\it NXB1} data, and
table \ref{stat_model}(b) shows their reproducibilities.
We can improve reproducibilities by employing the orbit filter.
Especially, the reproducibility of the XIS0 is 3 times better than 
that without orbit filter.

\begin{figure}[htbp]
  \begin{center}
    \FigureFile(80.0mm,58.0mm){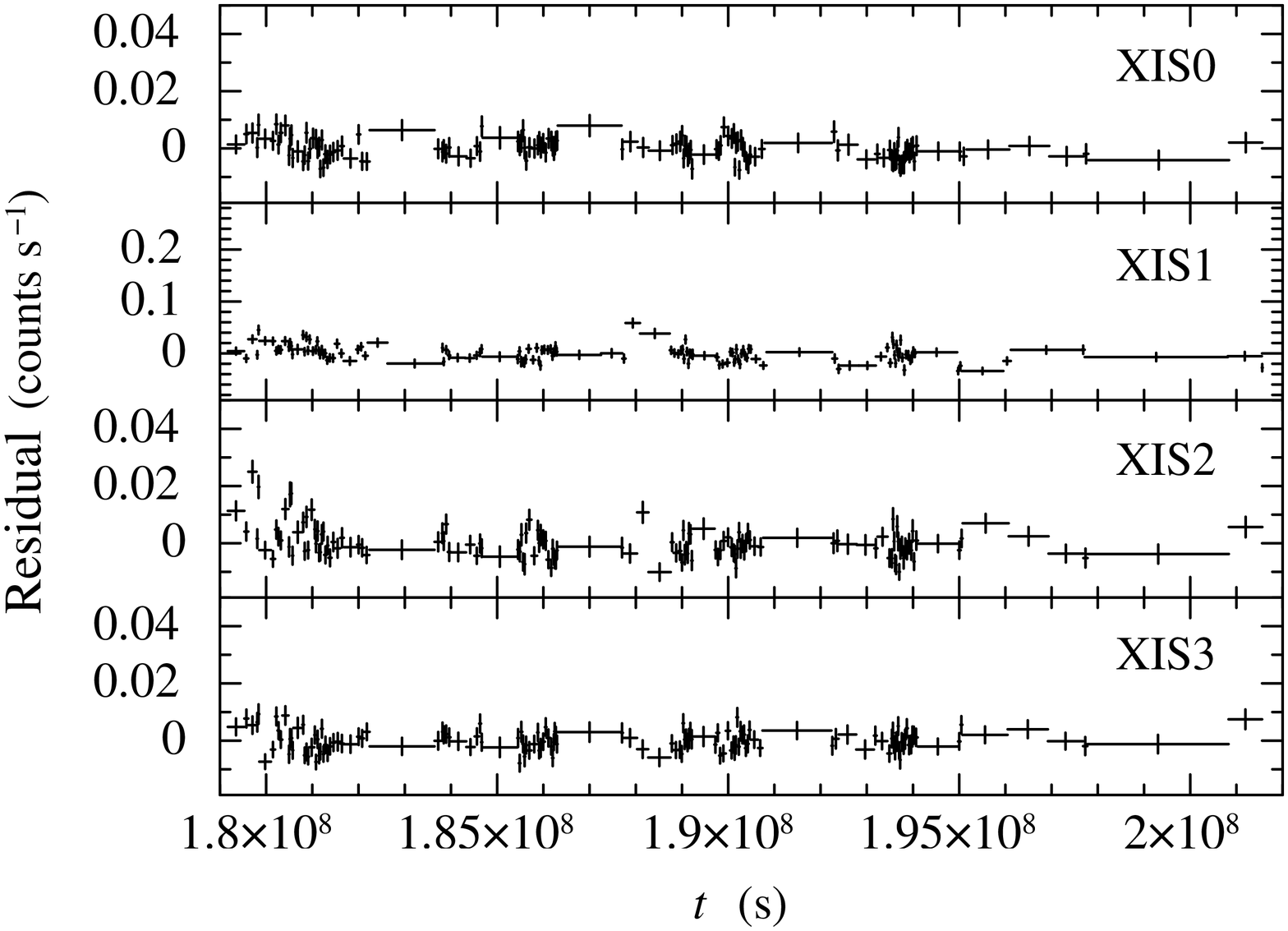}
  \end{center}
  \caption{Light curve of $\Delta C_{{\rm PIN-UD}}$ 
in the XIS0, 1, 2, and 3.
The residuals are obtained from the {\it NXB1} data 
in 5-12\,keV energy band.}
    \label{of_lc}
\end{figure}

The NXB model of XIS2 with the orbit filter 
applied has almost the same level of the reproducibility as
that without the orbit filter. 
In addition, the reproducibility 
of XIS2 is the worst among the XIS-FIs (XIS0, 2, 3) 
(table \ref{stat_model}(a) and (b)).
We therefore investigate the long-term variation of the NXB intensity.
Figure \ref{of_lc} shows the light curve of 
$\Delta C_{{\rm PIN-UD}}$ obtained from the {\it NXB1} data.
We found that $\Delta C_{{\rm PIN-UD}}$ during September 2005
had been higher than that after October 2005, 
especially in the XIS2.
We speculate that this is because the solar activity was particularly high
during September 2005.
The proton and solar X-ray intensities are continuously monitored 
by the Geostationary Operational Environmental Satellites 
(GOES)\footnote{The GOES data are available at 
$\langle$\,http://www.ngdc.noaa.gov/stp/GOES/goes.html\,$\rangle$}.
These intensities in September 2005
frequently exceeded 100 times
those of the normal state of the Sun.
We fitted the light curve after October 2005 with a linear function, 
finding that the NXB intensities of the XIS-FIs were 
constant with time within $\pm 6$\,\% per year.
On the other hand, the NXB intensity of XIS1 decreased
with a gradient of $(-7.8 \pm 5.8)$\,\% per year 
(90\,\% confidence level).
However, since the gradient is small, 
we continue to apply the same method of modeling 
as the XIS-FIs to XIS1.

We therefore exclude the NXB events during September 2005 
from the {\it NXB1} data (hereafter ``{\it NXB2}'') and
independently evaluate their reproducibilities for each XIS.
Figure \ref{of_distri}(b) shows the distribution 
of $\Delta C_{{\rm PIN-UD}}$ obtained from the {\it NXB2} data, and
table \ref{stat_model}(c) shows their reproducibilities.
Better reproducibilities than the
unfiltered NXB data can be obtained for all the XIS.
In addition, we found that 
the NXB model with the PIN-UD has better reproducibility 
than that with the {\it COR2}.
The total exposure time of this {\it NXB2} data is $\sim 560$\,ks,
but for the PIN-UD model, the exposure time is $\sim 550$\,ks.
Since the exposure time is long enough, 
excluding the data during September 2005 is not a serious problem 
for the observations after October 2005.

\begin{table}[htbp]
  \caption{Reproducibilities of two kinds of the NXB models
which are calculated for the 5\,ks NXB data in the 5-12\,keV energy band.}
  \label{stat_model}
  \begin{center}
    \begin{tabular}{l l c c c c c}
      \hline \hline
NXB & Sensor & {\it COR2}\footnotemark[$*$] (\%) & PIN-UD\footnotemark[$*$] (\%) \\
      \hline
        (a) Unfiltered & XIS0 & $6.60 \pm 0.61$ & $8.12 \pm 0.68$ \\
        & XIS1 & $7.18 \pm 0.46$ & $5.57 \pm 0.39$ \\
        & XIS2 & $10.33 \pm 0.77$ & $7.85 \pm 0.68$ \\
        & XIS3 & $3.32 \pm 0.56$ & $3.42 \pm 0.57$ \\
      \hline
        (b) {\it NXB1} & XIS0 & $4.05 \pm 0.55$ & $2.67 \pm 0.53$ \\
        & XIS1 & $6.32 \pm 0.43$ & $4.81 \pm 0.36$ \\
        & XIS2 & $9.77 \pm 0.76$ & $8.49 \pm 0.72$ \\
        & XIS3 & $4.09 \pm 0.59$ & $4.11 \pm 0.60$ \\
      \hline
        (c) {\it NXB2} & XIS0 & $3.54 \pm 0.61$ & $2.79 \pm 0.60$ \\
        & XIS1 & $6.95 \pm 0.53$ & $4.36 \pm 0.39$ \\
        & XIS2 & $5.67 \pm 0.69$ & $3.96 \pm 0.64$ \\
        & XIS3 & $2.34 \pm 0.64$ & $3.82 \pm 0.68$ \\
      \hline
      \multicolumn{4}{@{}l@{}}{\hbox to 0pt{\parbox{180mm}{\footnotesize
\par\noindent 
\footnotemark[$*$] These values are normalized by the average count rate
shown in \\[-5pt] table \ref{stat_nxbdata}. 
Errors are $1 \sigma$ confidence level.
	  }\hss}}
    \end{tabular}
  \end{center}
\end{table}

\subsection{Reproducibility of the NXB for the 1-7\,keV band}

We evaluate the reproducibility of the NXB in the 1-7\,keV energy band
by the same method as that in section \ref{sec_repro}.
The CXB is dominant compared with the NXB in this energy band.
Table \ref{stat_2} shows the average count rate, 
statistical error, and reproducibility for the 5\,ks {\it NXB2} data.
The reproducibilities are as good as those for the 
5-12\,keV energy band (table \ref{stat_model}(c)), and 
the NXB model with the PIN-UD has better reproducibility 
than that with the {\it COR2}.

\begin{table}[htbp]
  \caption{Reproducibilities of the NXB models 
which are calculated by dividing the {\it NXB2} data into 5\,ks exposure bins
in the 1-7\,keV energy band.}
  \label{stat_2}
  \begin{center}
    \begin{tabular}{l c c c c c c c c}
      \hline \hline
Sensor & Average count rate & Statistical error\footnotemark[$*$] & \multicolumn{2}{c}{Reproducibility\footnotemark[$*$]} \\
 & $10^{-2}$ (cts s$^{-1}$) & (\%) & {\it COR2} (\%) & PIN-UD (\%) \\
      \hline
XIS0 & $4.163 \pm 0.027$ & $6.94 \pm 0.47$ & $5.06 \pm 0.74$ & $4.55 \pm 0.74$ \\
XIS1 & $7.321 \pm 0.036$ & $5.22 \pm 0.35$ & $7.55 \pm 0.71$ & $5.63 \pm 0.63$ \\
XIS2 & $3.871 \pm 0.026$ & $7.19 \pm 0.49$ & $7.31 \pm 0.84$ & $5.18 \pm 0.78$ \\
XIS3 & $3.475 \pm 0.025$ & $7.59 \pm 0.52$ & $6.34 \pm 0.84$ & $4.76 \pm 0.80$ \\
      \hline
      \multicolumn{5}{@{}l@{}}{\hbox to 0pt{\parbox{180mm}{\footnotesize
\par\noindent 
\footnotemark[$*$] These values are normalized by the average count rate.
Errors are $1 \sigma$ confidence level.
	  }\hss}}
    \end{tabular}
  \end{center}
\end{table}

\subsection{Reproducibility with longer exposure data}

We have so far calculated the NXB reproducibility by dividing 
the NXB data into each 5\,ks exposure bins.
Here we examine the NXB reproducibility for a longer 
exposure time of 50\,ks, typical for on-source observations.
Each 50\,ks NXB exposure typically spans a few weeks.
Table \ref{stat_50ks} shows the reproducibilities of the NXB models 
for the exposure time of 50\,ks 
in the energy bands of 1-7\,keV and 5-12\,keV.
The reproducibilities of the 50\,ks NXB models are 
improved from those for the 5\,ks NXB models.
This is because fluctuations of the residuals are smoothed by 
integrating for a long time.

\begin{table}[htbp]
  \caption{Reproducibilities of the NXB models which 
are calculated by dividing the {\it NXB2} data into 50\,ks exposure bins.}
  \label{stat_50ks}
  \begin{center}
    \begin{tabular}{l l c c c c c c}
      \hline \hline
Energy range & Sensor & Statistical error\footnotemark[$*$] & \multicolumn{2}{c}{Reproducibility\footnotemark[$*$]} \\
(keV) & & (\%) & {\it COR2} (\%) & PIN-UD (\%) \\
      \hline
$1 - 7$ & XIS0 & $2.29 \pm 0.54$ & $1.89 \pm 0.84$ & $2.02 \pm 0.90$ \\
 & XIS1 & $1.72 \pm 0.39$ & $2.61 \pm 0.80$ & $2.70 \pm 0.81$ \\
 & XIS2 & $2.37 \pm 0.56$ & $1.73 \pm 0.84$ & $0.31 \pm 0.79$ \\
 & XIS3 & $2.50 \pm 0.59$ & $2.08 \pm 0.92$ & $1.20 \pm 0.88$ \\
      \hline
$5 - 12$ & XIS0 & $1.96 \pm 0.46$ & $1.03 \pm 0.66$ & $1.89 \pm 0.79$ \\
 & XIS1 & $0.85 \pm 0.19$ & $2.98 \pm 0.72$ & $2.36 \pm 0.59$ \\
 & XIS2 & $1.98 \pm 0.47$ & $1.87 \pm 0.75$ & $1.20 \pm 0.72$ \\
 & XIS3 & $2.14 \pm 0.50$ & $1.51 \pm 0.75$ & $0.40 \pm 0.72$ \\
      \hline
      \multicolumn{5}{@{}l@{}}{\hbox to 0pt{\parbox{180mm}{\footnotesize
\par\noindent 
\footnotemark[$*$] These values are normalized by the average count rate
in the energy bands of the \\[-5pt]
1-7\,keV (table \ref{stat_2}) or the 5-12\,keV (table \ref{stat_nxbdata}).
Errors are $1 \sigma$ confidence level.
	  }\hss}}
    \end{tabular}
  \end{center}
\end{table}

\section{Subtraction of the NXB from on-source observation}\label{subtract}

In this section, we consider the practical manner 
of how to subtract the NXB for on-source science observations.
First, since the intensity of the NXB is not uniform over the CCD chip
(\cite{yamaguchi_spie}), the NXB spectrum needs to be extracted 
from the same region as the source spectrum in detector (DET)
coordinates (\cite{ishisaki}).
We can extract the NXB spectra sorted by the cut-off-rigidity for a given
region defined in the DET coordinate with the {\it mk\_corsorted\_spec}
(same applies to the {\it mk\_corsorted\_spec\_v1.0.pl}).
Next, the sorted NXB spectra are summed up with 
appropriate weights calculated for the on-source observation
using the {\it mk\_corweighted\_bgd}
(same applies to the {\it mk\_corweighted\_bgd\_v1.1.pl}).
The summed-up spectrum is the NXB model to be subtracted from 
on-source spectra.

\begin{figure}[htbp]
  \begin{center}
    \FigureFile(80.0mm,62.5mm){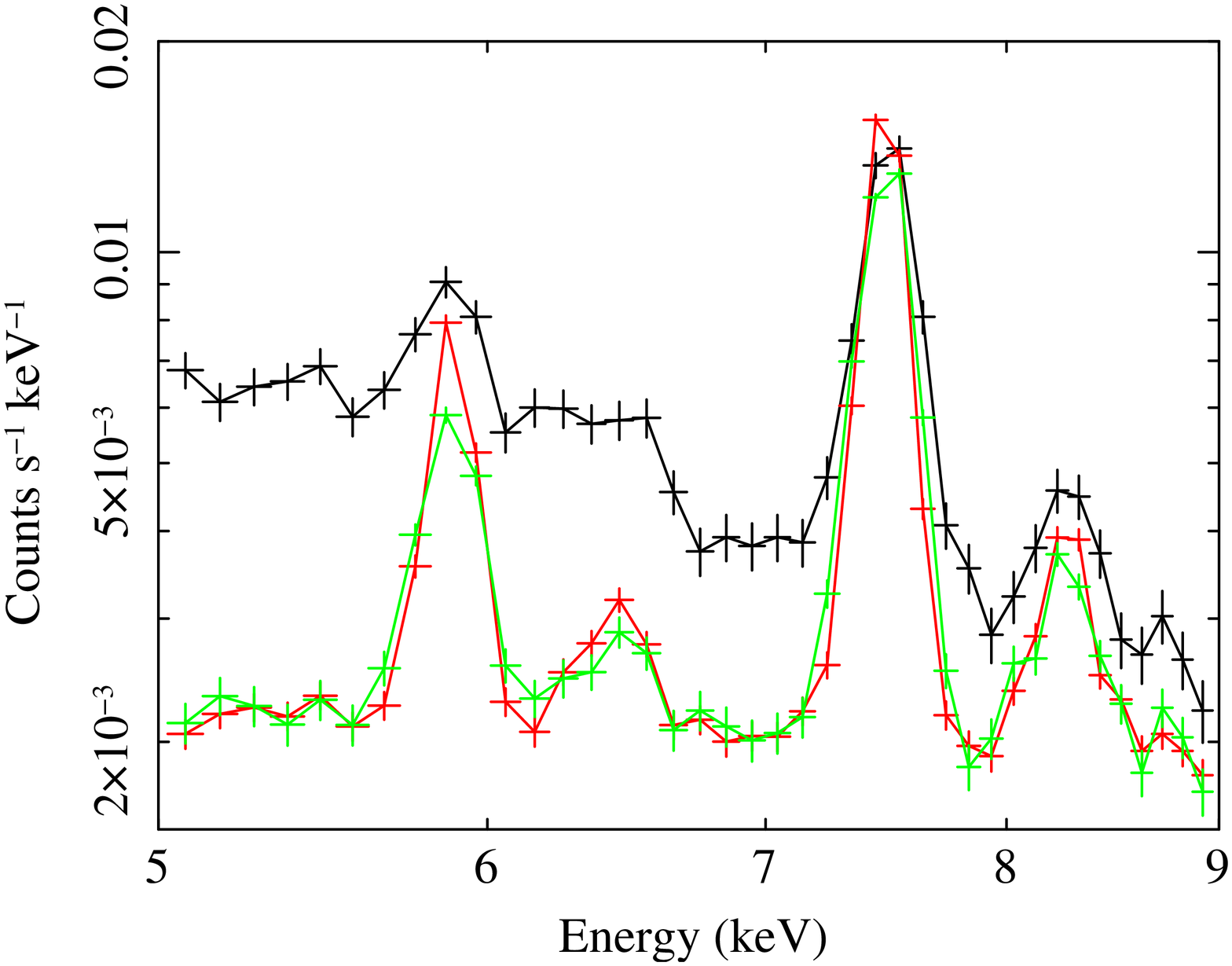}
  \end{center}
  \caption{(black) On-source observation spectrum.
(red) The NXB model spectrum without the correction for the emission 
line components.
(green) The NXB model spectrum with the correction for the emission 
line components.
These are averaged spectra of the XIS-FIs.}
    \label{spec_comp}
\end{figure}

\begin{figure}[htbp]
  \begin{center}
    \FigureFile(160.0mm,55.7mm){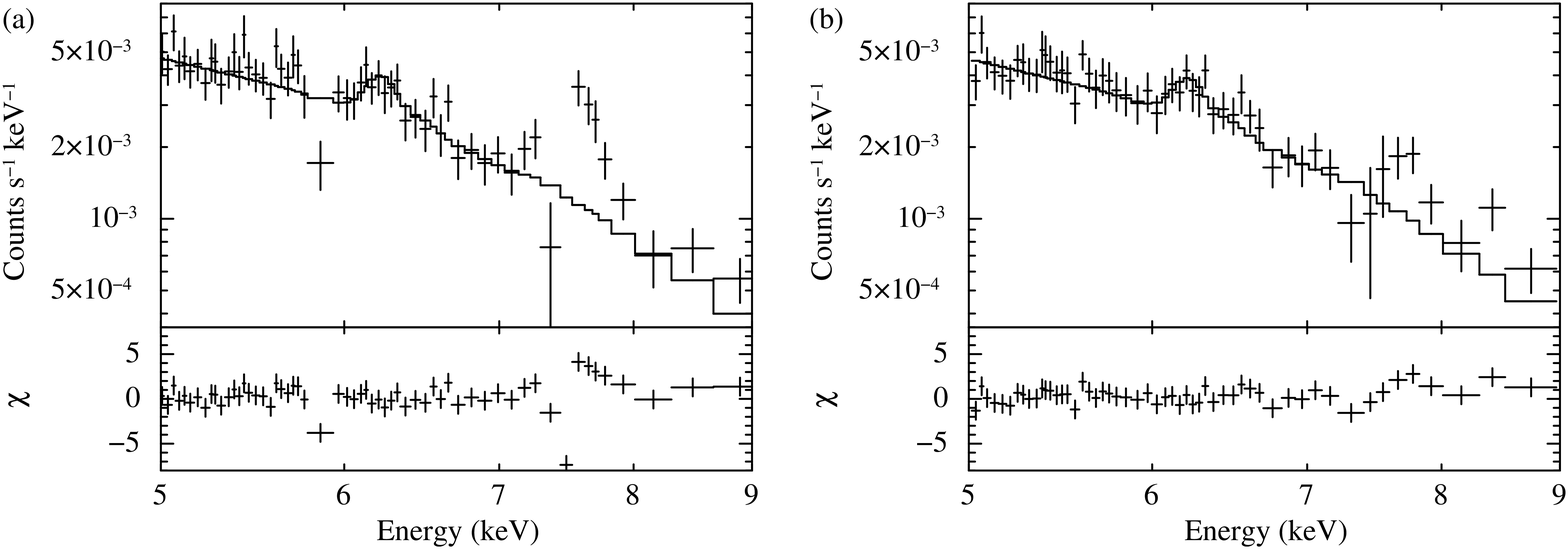}
  \end{center}
  \caption{On-source observation with or without the 
correction for the emission line components.
(a) The spectrum from which the NXB model spectrum is subtracted 
without the correction for the emission line components.
(b) The spectrum from which the NXB model spectrum is subtracted 
with the correction for the emission line components.
These are averaged spectra of the XIS-FIs.}
    \label{spec_bef}
\end{figure}

One of the problems in this procedure is the presence of 
emission line components
in the NXB spectra.
These components are time-dependent; the energy resolution 
of the XIS degrades with time (\cite{koyama}), 
and the intensities of the Mn-K emission lines 
decrease with the half life of $^{55}$Fe, 2.73 years.
Since the NXB data contained in the NXB database 
are made from the NXB events between September 
2005 and May 2006 and this time dependence is not taken into account, 
the emission line components 
in the NXB spectra cannot be reproduced correctly 
for a given on-source observation. 
This problem becomes prominent for the observations after June 2006.
Figure \ref{spec_comp} shows an example of a raw on-source 
spectrum and the NXB model spectrum which 
we have described above.
These are the averaged spectra of the XIS-FIs.
Figure \ref{spec_bef}(a) shows the on-source spectrum 
from which the NXB model spectrum 
is subtracted, black minus red line shown in Fig. \ref{spec_comp}.
This on-source spectrum is obtained in the observation of the link 
region between the galaxy clusters A 399 and A 401 taken during 
August 19-22 2006 
with an exposure time of 150\,ks (observation ID is 801020010.
For details, see Fujita et al. 2007).
The model in Fig. \ref{spec_bef}(a) is 
a single thermal model (APEC in XSPEC)
plus power-law model. The thermal model represents the intracluster 
medium, and the power-law model represents the CXB.
These are same models used by Fujita et al. (2007).
Significant residuals are visible at the energies of 
Mn-K$\alpha$ (5.9\,keV) and Ni-K$\alpha$ (7.4\,keV)
in Fig. \ref{spec_bef}(a).
We have therefore developed a way 
to deal with these emission line components in the NXB spectra, as follows: 

\begin{enumerate}

\item The NXB model spectrum is constructed with the {\it COR2}
or the PIN-UD using the method described in subsection \ref{howto_model}.

\item The line components Mn-K$\alpha$, Mn-K$\beta$, 
Ni-K$\alpha$, Ni-K$\beta$, and Au-L$\alpha$, 
in the NXB model spectrum are fitted 
with the redistribution matrix file (RMF) %response
for August 2005 observations, 
at which point the degradation of the energy resolution was negligible.
For example, in the energy range of 
5.5-7.0\,keV where 
there are Mn-K$\alpha$ and Mn-K$\beta$ lines,
the spectrum is fitted with two Gaussians plus a power-law continuum. 
We have set the line widths of the 
two Gaussian components as free parameters. 
The emission lines of Ni-K$\alpha$, Ni-K$\beta$, and
Au-L$\alpha$ are similarly fitted.

\item We simulate the spectrum of the five Gaussian components 
using the {\it fakeit} command in XSPEC, 
using the fitting parameters determined in step 2.

\item The spectrum created in the step 3 is subtracted from 
the NXB model spectrum from step 1.
This should correspond to the NXB 
continuum spectrum from which the five Gaussian line components
are removed.

\item We add the simulated line components to the NXB continuum spectrum 
created in step 4, using the {\it fakeit} command.
To take into account the degradation of energy resolution, 
this simulation needs to be done with the RMF calculated 
for the epoch of the on-source observation by using 
the {\it xisrmfgen} command in FTOOLS.
{\it Xisrmfgen} is a response generator for the Suzaku XIS.
The normalizations and line center energies of the five Gaussian components 
are fixed with those obtained in step 2, though 
radioactive decay of $^{55}$Fe is taken into account.
The intrinsic widths of these lines are fixed to be zero.
We then get the NXB model spectrum in which degradation in 
the energy resolution and the $^{55}$Fe decay are
taken into account.

\end{enumerate}

The NXB model spectrum with the correction for the emission line components
is shown as a green line in Fig. \ref{spec_comp}.
The line widths of this spectrum are larger than those of the NXB model
spectrum without the correction (red line shown in Fig. \ref{spec_comp}).
Additionally, the intensities of Mn-K lines decrease with the correction.
Figure \ref{spec_bef}(b) shows the source spectrum from which 
the NXB model with the correction for the emission line components
is subtracted.
The model in Fig. \ref{spec_bef}(b) is the same as that in 
Fig. \ref{spec_bef}(a).
This correction can reduce the residuals in 
the energy bands including Mn-K$\alpha$ and Ni-K$\alpha$, 
improving the reduced $\chi^2$ from 1.77 to 1.16, in this case.

\section{Discussion}

\subsection{Case study: 100\,ks exposure}

In this section, we demonstrate to what extent 
the uncertainty of the source intensity depends on the NXB reproducibility.
We assume an extended
source over the XIS FOV whose surface brightness is comparable to that
of the CXB. We also assume an exposure time of 100\,ks, 
typical for this type of extended source. 
Such observation typically spans two days, 
corresponding to an NTE exposure of 5\,ks in our database.
We will concentrate on the energy band of 5-12\,keV of the XIS-FIs.
We find the count rate of the NXB, $I_{NXB}$, to be $5.0 \times 10^{-2}$ 
cts s$^{-1}$ 
and the reproducibility, $\Delta I_{NXB}$, to be $1.8 \times 10^{-3}$ 
cts s$^{-1}$
(3.5\,\% of the $I_{NXB}$) based on table \ref{stat_model}(c).
Kushino et al. (2002) measured 
a CXB power-law photon index of 
1.412 and flux of $6.38 \times 10^{-8}
{\rm \ erg\ cm^{-2}\ s^{-1}\ sr^{-1}}$.
They also estimated the spatial fluctuation of the CXB flux 
to be 6.5\,\% ($1 \sigma$) by analyzing the ASCA GIS data. 
Employing these values, we can evaluate the expected count
rate for the XIS-FIs, $I_{CXB}$, to be $9.7 \times 10^{-3}$ cts s$^{-1}$.
We assume that the spatial fluctuation follows to the Poisson statistics
of the number of sources in the FOV. 
Then, we can calculate the spatial fluctuation of the CXB 
for the XIS FOV to be $6.5 \times \sqrt{0.4/0.088} = 14\,\%$, 
since the FOVs of the ASCA GIS and the XIS are $0.4$ deg$^2$ and
$0.088$ deg$^2$, respectively. 
Employing the exposure of 100ks,
we can estimate the CXB fluctuation in the XIS, $\Delta I_{CXB}$, 
to be $1.4 \times 10^{-3}$ cts s$^{-1}$.  
We should note that $\Delta I_{CXB}$ is
comparable to $\Delta I_{NXB}$ in our case.

When we obtain the count rate of the raw data, $I_{raw}$, we will subtract
the NXB and CXB from it in order to evaluate the source count rate, $I_{src}$.
Since we assume $I_{src}$ to be comparable to $I_{CXB}$, $I_{raw}$ is 
$(5.0 + 0.97 + 0.97) \times 10^{-2} = 6.9 \times 10^{-2}$ cts s$^{-1}$ 
and its statistical error, $\Delta I_{raw}$, is $8.3 \times 10^{-4}$ 
cts s$^{-1}$. 
We will obtain $I_{src}$ by subtracting $(I_{CXB} + I_{NXB})$ from $I_{raw}$. 
We can calculate the error
of $I_{src}$, $\Delta I_{src}$, to be 
$\sqrt{ \Delta I_{NXB}^2 + \Delta I_{CXB}^2 + \Delta I_{raw}^2} 
= 2.4 \times 10^{-3}$ cts s$^{-1}$.  
$\Delta I_{NXB}$ and $\Delta I_{CXB}$ contribute almost equally to
$\Delta I_{src}$, while the contribution of the
statistical error $\Delta I_{raw}$  is smaller than these two.
Similarly, we evaluate the NXB reproducibility of the XIS-FIs in
the 1-7\,keV energy band by using the same method as employed in
the 5-12\,keV band. 
The calculations show that $\Delta I_{NXB} = 1.7 \times 10^{-3}$,
$\Delta I_{CXB} = 8.1 \times 10^{-3}$, and 
$\Delta I_{src} = 1.3 \times 10^{-3}$ (cts s$^{-1}$).
We should note that $I_{NXB}$ does not change
so much while $I_{CXB}$ is 6 times bigger in the 1-7\,keV band than that in
the 5-12 keV band. 
Therefore, $\Delta I_{src}$ is mainly determined by $\Delta I_{CXB}$
rather than by $\Delta I_{NXB}$.

\subsection{Case study: filtering the data to reduce the error of source count rate}

We have considered the case of a 100\,ks exposure in the subsection above.
Depending on the exposure time and the count rate of the source,
the contribution of the statistical error, $\Delta I_{raw}$, to
$\Delta I_{src}$ is minor.
$\Delta I_{src}$ can be reduced
in some conditions by filtering the data as shown below.

We consider a specific example in which the data are
filtered with the condition of PIN-${\rm UD} < 225$ cts s$^{-1}$. 
We call this as the ``PIN-UD filter''. 
Using the PIN-UD filter,
the exposure time reduces to 85\,\% of that without the filter.
The NXB count rate and reproducibility in the 5-12\,keV band of the XIS-FIs
are $4.6 \times 10^{-2}$ cts s$^{-1}$ ($I'_{NXB}$) and 3.2\,\% 
(normalized by $I'_{NXB}$), respectively.
These values are based on the {\it NXB2} data,
and this reproducibility is obtained for each 5\,ks exposure.

We will derive
the condition that the PIN-UD filter provides a smaller
$\Delta I_{src}$ than that without the PIN-UD filter
for a given on-source observation.
We hereafter refer to the error of source count rate and 
the NXB reproducibility with the PIN-UD filter as $\Delta {I'}_{src}$
and $\Delta {I'}_{NXB}$, respectively.
On the other hand, those without the PIN-UD filter are
newly defined as $\Delta I_{src}$ and $\Delta I_{NXB}$, respectively.
$\Delta {I'}_{src}$ is expressed as below,
\begin{eqnarray}
\Delta {I'}_{src}^2 &=& \Delta {I'}_{NXB}^2 + \Delta I_{CXB}^2 
+ \Delta {I'}_{raw}^2 , \\
&=& (0.032 {I'}_{NXB})^2 + \Delta I_{CXB}^2 + 
\frac{I_{CXB} + {I'}_{NXB} + I_{src}}{0.85T} \ , 
\label{eq_err_pinf}
\end{eqnarray}
where $\Delta {I'}_{raw}$ is the statistical error of the count rate of 
the raw data with the PIN-UD filter. 
$T$ is the exposure time of the on-source observation.
On the other hand, $\Delta I_{src}$ is expressed as below,
\begin{eqnarray}
\Delta I_{src}^2 = (0.035 I_{NXB})^2 + \Delta I_{CXB}^2 + 
\frac{I_{CXB} + I_{NXB} + I_{src}}{T} \ .
\label{eq_err_org}
\end{eqnarray}
$I_{src}$, $I_{CXB}$, and $\Delta I_{CXB}$ are not altered by 
the PIN-UD filter.
To obtain a value for $\Delta {I'}_{src}$ 
which is smaller than that for $\Delta I_{src}$,
the condition is expressed as below,
\begin{eqnarray}
(0.032 I'_{NXB})^2 + \frac{I_{CXB} + I'_{NXB} + I_{src}}{0.85T}
< (0.035 I_{NXB})^2 + \frac{I_{CXB} + I_{NXB} + I_{src}}{T} \ , \\
I_{src} < 5.1 \times 10^{-3} \biggl( \frac{T}{1\,{\rm ks}} \biggr)
- 0.033 \ \ {\rm cts\ s^{-1}} \ .
\label{eq_stat}
\end{eqnarray}
The term of $\Delta I_{CXB}^2$ vanishes from both sides of the above 
inequality.

In the case that the exposure time of the on-source observation is 100\,ks,
the PIN-UD filter is effective for a diffuse source whose count rate
is less than $0.48$ cts s$^{-1}$.
This count rate corresponds to 49 times that of the CXB.
On the other hand, if the exposure time is 20\,ks,
the PIN-UD filter is effective only for a source whose count rate
is lower than 7.1 times of that of the CXB.

\section{Summary}

We have constructed the NXB database by collecting the XIS events
of the NTE.
The NXB database, accompanied with EHK files, and two software tools,
{\it mk\_corsorted\_spec\_v1.0.pl} and {\it mk\_corweighted\_bgd\_v1.1.pl}, 
are now accessible
via the Suzaku web page at ISAS/JAXA and GSFC/NASA.
Since the XIS NXB depends on the cut-off-rigidity 
in orbit or on the PIN-UD count rate,
we need to equalize the distributions of these parameters for the on-source
observations and for the NTE observation so that we can actually 
subtract the NXB.
We have examined two modeling parameters to model the NXB, 
the {\it COR2} and the PIN-UD.
We find large deviation of the NXB count rate from that expected
from both models when Suzaku passes through high altitude 
and high latitude regions.
Excluding those data, the NXB reproducibilities are significantly improved.
Similarly, excluding the data taken in September 2005, the
reproducibility for the XIS2 is improved.
Our results show that the NXB model sorted by the PIN-UD has 
better reproducibility than that by the {\it COR2}.
Using the NXB data in 5 ks exposure bins, 
the reproducibility obtained with the PIN-UD model is 
4.55-5.63\,\% for each XIS NXB in the 1-7\,keV band and 
2.79-4.36\,\% for each XIS NXB in the 5-12\,keV band.
This NXB reproducibility in 5-12\,keV, $1.8 \times 10^{-3}$ cts s$^{-1}$, 
is comparable to the 
spatial fluctuation of the CXB for the XIS FOV, 
$1.4 \times 10^{-3}$ cts s$^{-1}$.
The NXB reproducibility and the spatial fluctuation of the CXB
are evaluated to be $1.7 \times 10^{-3}$ cts s$^{-1}$ 
and $8.1 \times 10^{-3}$ cts s$^{-1}$, respectively, in the 1-7 keV band.
Depending on the exposure time and the count rate of the source, 
the statistical error of the raw data is much smaller than
the NXB reproducibility.
In such a case, 
the error of the source count rate can be reduced
by excluding the data with high NXB count rate
(e.g. filtering with PIN-UD $< 225$ cts s$^{-1}$).

\bigskip

The authors wish to thank all the XIS team members for their support, 
help, and useful information.
This work is partly supported by a Grant-in-Aid for Scientific
Research by the Ministry of Education, Culture, Sports, Science and
Technology (16002004 and 1910350).  This study is also carried out as part of
the 21st Century COE Program, \lq{\it Towards a new basic science:
depth and synthesis}\rq.  N. T. is supported by JSPS Research Fellowship
for Young Scientists.

\appendix
\section{New and old map for the cut-off-rigidity}\label{oldCOR}

\begin{figure}[htbp]
  \begin{center}
    \FigureFile(160.0mm,64.5mm){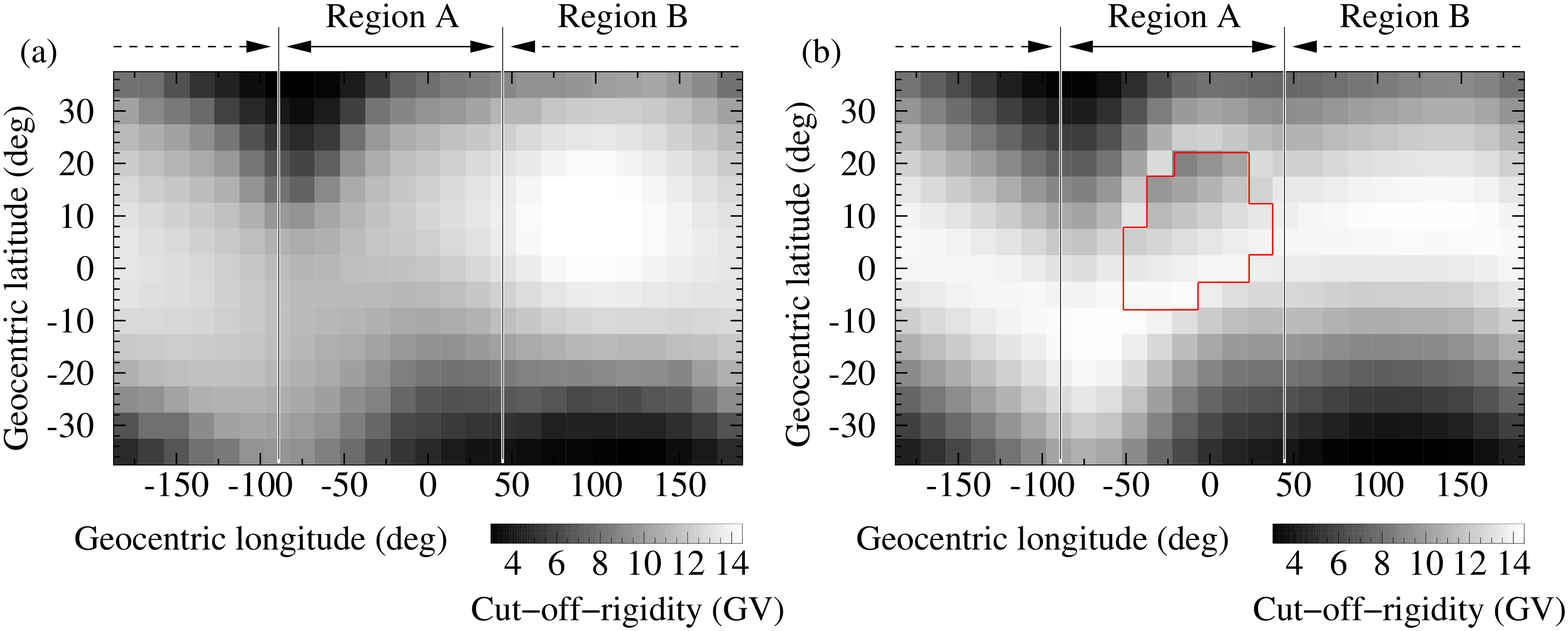}
  \end{center}
  \caption{Maps of (a) the {\it COR} and (b) the {\it COR2}.
The region A is $-90^{\circ} < {\rm latitude} < 45^{\circ}$, and
the region B encompasses the remaining latitude range.}
  \label{cor_map}
\end{figure}

The cut-off-rigidity values have been calculated from 
the orbital position of Suzaku
using a cut-off-rigidity map shown in Fig. \ref{cor_map}(a).
However, the map assumes charged particles originating from the
zenith direction at an altitude of 500\,km, and it uses 
an international geomagnetic reference field for 1975.
This cut-off-rigidity definition (hereafter ``{\it COR}'') 
is out of date and inaccurate. 
We therefore define a new cut-off-rigidity map 
based on the recent cut-off-rigidity database.

The new cut-off-rigidity map is calculated by using 
corrected geomagnetic (CGM) coordinates.
The CGM coordinates are useful to study geophysical phenomena affected
by the Earth's magnetic field and are provided at the
web service by NASA\footnote{The service
is available at 
$\langle$\,http://modelweb.gsfc.nasa.gov/models/cgm/t96.html\,$\rangle$}
(\cite{cgm2}).
To calculate the CGM coordinates, several parameters are required.
We set these parameters as follows; altitude is 570km, date is
2006/01/01 00:00:00, and
default parameters are employed for the
solar wind (${\rm Den}=3$, ${\rm Vel}=400$,
${\rm BY}=5$, ${\rm BZ}=-6$, and ${\rm Dst}=-30$).
Then, the new cut-off-rigidity value, $R_C$, 
is calculated as follows,
\begin{eqnarray}
R_C = 14.5 \frac{\cos^2\theta}{r^2} \hspace{3mm} {\rm GV},
\label{eq1}
\end{eqnarray}
where $\theta$ is the latitude in CGM coordinates.
$r$ is the distance from the center of earth's magnetism, and 
the value of $r$ is normalized by the radius of the earth.
We call this cut-off-rigidity as ``{\it COR2}''.
Figure \ref{cor_map}(b) shows the {\it COR2} map.
In the red box region shown in Fig. \ref{cor_map}(b),
since the CGM cannot be obtained due to the local magnetic structure, 
we use geomagnetic latitude obtained with dipole approximation
in place of the CGM.

The cut-off-rigidity value for each event can be determined from 
the EHK file associated with each observation.
The EHK files before revision 2.0 processing contain only the {\it COR},
while those after revision 2.0 contain both 
the {\it COR} and the {\it COR2}.
These {\it COR} and {\it COR2} values are calculated using the cut-off-rigidity
maps of {\it rigidity\_20000101.fits} and {\it rigidity\_20060421.fits} in
the generic area of the calibration database (CALDB).

We evaluate the reproducibility of the NXB model with the {\it COR}
by the same way as that with the {\it COR2} or the PIN-UD in main text.
The bin ranges of the {\it COR} to sort the NXB data and on-source data
are the same as the {\it COR2} as shown in table \ref{range}.
Table \ref{stat_cor} shows the reproducibility of the NXB model 
with the {\it COR} in the energy bands of 1-7\,keV and 5-12\,keV.
The reproducibility is calculated by dividing the {\it NXB2} data
into 5\,ks or 50\,ks exposure bins.
We found that the {\it COR} has the worst reproducibility 
among the three kinds of the NXB models.

\begin{figure}[htbp]
  \begin{center}
    \FigureFile(160.0mm,63.1mm){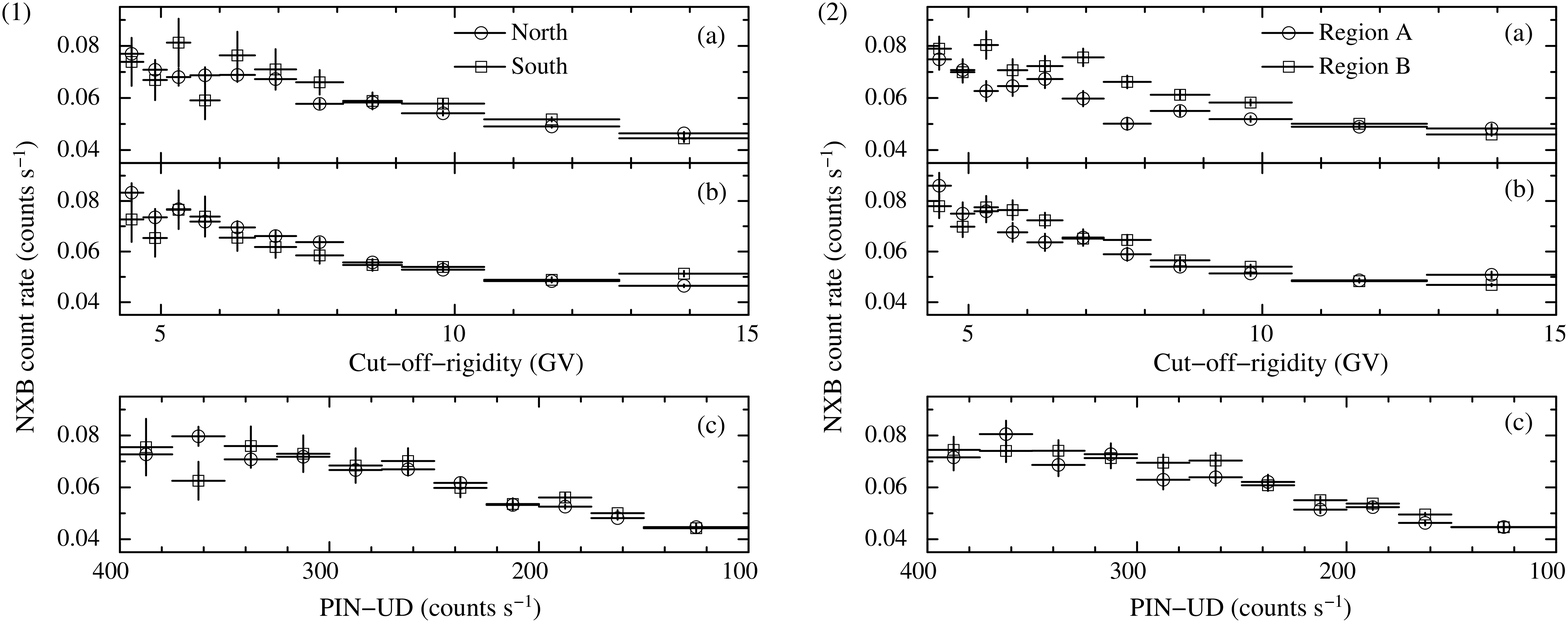}
  \end{center}
  \caption{NXB of the XIS0 for given values of (a) {\it COR}, 
(b) {\it COR2}, and (c) PIN-UD. 
(1) shows the NXB count rates for the northern
and southern hemisphere separately. 
Open circles are for the north hemisphere and open squares for
the south hemisphere.
(2) shows the count rates for
region A and region B (see Fig. \ref{cor_map}).
Open circles are for region A and open squares for region B.}
  \label{dep_cor2}
\end{figure}

We searched for the location where 
the {\it COR} does not perfectly reproduce
the XIS NXB.
Figure \ref{dep_cor2}(1) shows the average count rate of the XIS0 NXB 
in the 5-12\,keV energy band 
as a function of the three modeling parameters, in the northern hemisphere and
in the southern hemisphere, displayed separately.
This count rate is obtained from the {\it NXB2} data.
For the three modeling parameters, there is not 
a large difference between the NXB count rate 
in the northern hemisphere and that in the southern hemisphere.
On the other hand, if we take the data from two different longitude regions,
region A and region B in Fig. \ref{cor_map}, we obtain Fig. \ref{dep_cor2}(2).
There is a significant difference between the NXB count rate
in region A and that in region B for a given value of the {\it COR}.
This is one reason the {\it COR} gives the worst reproducibility.

\begin{table}[htbp]
  \caption{Reproducibility of the NXB model with the {\it COR} 
in the energy bands of 1-7\,keV and 5-12\,keV.}
  \label{stat_cor}
  \begin{center}
    \begin{tabular}{l c c c c c c c}
      \hline \hline
Sensor & \multicolumn{5}{c}{Reproducibility of {\it COR} (\%)} \\
 & \multicolumn{2}{c}{5\,ks exposure} & & \multicolumn{2}{c}{50\,ks exposure} \\
\cline{2-3} \cline{5-6}
 & 1-7\,keV\footnotemark[$*$] & 5-12\,keV\footnotemark[$\dagger$] & & 1-7\,keV\footnotemark[$*$] & 5-12\,keV\footnotemark[$\dagger$] \\
      \hline
XIS0 & $5.73 \pm 0.76$ & $4.46 \pm 0.64$ & & $2.28 \pm 0.88$ & $1.60 \pm 0.72$ \\
XIS1 & $7.97 \pm 0.73$ & $6.60 \pm 0.51$ & & $3.42 \pm 0.94$ & $3.12 \pm 0.75$ \\
XIS2 & $7.63 \pm 0.85$ & $5.60 \pm 0.68$ & & $1.70 \pm 0.84$ & $2.02 \pm 0.77$ \\
XIS3 & $6.58 \pm 0.85$ & $3.20 \pm 0.65$ & & $2.14 \pm 0.92$ & $1.15 \pm 0.72$ \\
      \hline
\multicolumn{5}{@{}l@{}}{\hbox to 0pt{\parbox{180mm}{\footnotesize
\par\noindent 
\footnotemark[$*$] These values are normalized by 
the average count rates shown in table \ref{stat_2}.\\[-5pt]
\footnotemark[$\dagger$] These values are normalized by 
the average count rates shown in table \ref{stat_nxbdata}.
	  }\hss}}
    \end{tabular}
  \end{center}
\end{table}

\section{Errors of the statistical parameters}\label{calc_stat}

We discuss the errors of $\sigma_c$, $\sigma_{sta,c}$, 
$\sigma_{sys,c}$, and $\sigma_{sys,\Delta c}$
as defined in section 3.2.
If the distribution of $C_j$ follows a Gaussian distribution whose
average and standard deviation are $\mu$ and $\sigma$ respectively,
then $(n-1) \sigma_c^2 / \sigma^2 \equiv X$ follows a $\chi^2$ distribution 
with $(n-1)$ degrees of freedom.
The expected value and variance of $X$ are $(n-1)$ and
$2(n-1)$, respectively.
Although the distribution of the statistical error of $X$ does not 
correctly follow a Gaussian distribution, 
the statistical error can be approximated with $\sqrt{2(n-1)}$.
Thus, the statistical error of $\sigma_c^2$ is 
$\sigma^2 \sqrt{2/(n-1)}$.
However, since $\sigma$ is a standard deviation of the parent population and
can not be obtained, we approximate that $\sigma$ equals $\sigma_c$.
As a result, the statistical error of $\sigma_c^2$ is 
$\sigma_c^2 \sqrt{2/(n-1)}$.
The statistical error of $\sigma_c$ ($\Delta \sigma_c$) is 
expressed by using the principle of error propagation as follows,
\begin{eqnarray}
\Delta \sigma_c = \frac{\sigma_c}{\sqrt{2(n-1)}} \ .
\label{a2-5}
\end{eqnarray}
In the same way, the statistical error of $\sigma_{sta,c}$ 
($\Delta \sigma_{sta,c}$) is obtained as follows,
\begin{eqnarray}
\Delta \sigma_{sta,c} = \frac{\sigma_{sta,c}}{\sqrt{2(n-1)}} \ .
\label{a2-6}
\end{eqnarray}
The statistical errors of $\sigma_{sys,c}$ ($\Delta \sigma_{sys,c}$) 
is expressed by using the principle of error propagation as follows,
\begin{eqnarray}
\Delta \sigma_{sys,c} = \frac{1}{\sigma_{sys,c} \sqrt{2(n-1)}} 
\sqrt{ \sigma_c^4 + \sigma_{sta,c}^4 } \ .
\label{a2-7}
\end{eqnarray}
In the same way as $\Delta \sigma_{sys,c}$, 
the statistical error of $\sigma_{sys,\Delta c}$ 
($\Delta \sigma_{sys,\Delta c}$) is expressed as follows 
\begin{eqnarray}
\Delta \sigma_{sys,\Delta c} = \frac{1}{\sigma_{sys,\Delta c} \sqrt{2(n-1)}} 
\sqrt{ \sigma_{\Delta c}^4 + \sigma_{sta,\Delta c}^4 } \ .
\end{eqnarray}

%%%
% See the manual for the detail.
%%%

\end{document}